# Alcohol induced surface charging of colloidal quantum dots for controllable electrophoretic deposition processing

Jiaming Su[1], Kai Gu[1], Qingchen Wang[1], Kaiying Min[1], Zhiyuan Gao[1], Haizheng Zhong[1]*.

[1]MIIT Key Laboratory for Low-Dimensional Quantum Structure and Devices, School of Materials Science and Engineering, Beijing Institute of Technology, Beijing 100081

*Corresponding authors. E-mail addresses: hzzhong@bit.edu.cn

**Abstract:** In this work, we report an alcohol-induced surface charging route of colloidal QDs to achieve controllable electrophoretic deposition processing. By adding a fixed amounts of alcohols into a preformed quantum dots solution in non-polar solvents, the colloidal quantum dots can be positively charged, and then deposited on negative electrode under applied electric field. The surface charging of PbSe quantum dots was investigated by zeta potential, nuclear magnetic resonance, Fourier transform infrared spectroscopy, and discrete Fourier transform calculations. It was found that the zeta potential of oleate acid capped PbSe QDs increases from +1.6 mV to +13.4 mV with the amount of alcohol solvent increasing. The alcohol-induced zeta potential increasing can be explained to the electron cloud shift of active hydrogen mediated by intermolecular hydrogen bonds between carboxy acid and alcohol. Considering the influence of surface charging of quantum dots on their dispersibility, we describe the microscopic mechanism of alcohol-induced electrophoretic deposition processing. Furthermore, we developed a size-selective separation protocol by controlling alcohol-induced electrophoretic deposition processing.

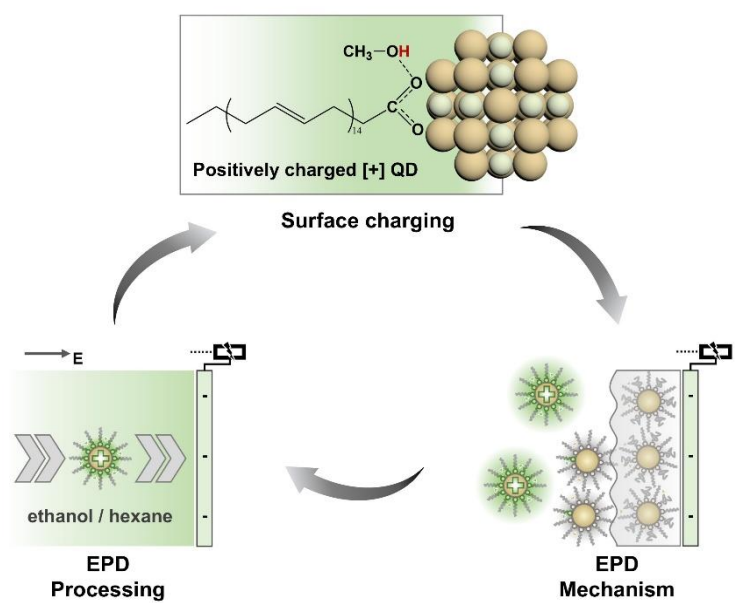

## Introduction

Colloidal quantum dots (QDs) are tiny nanocrystals with organic ligand on their surface, which have been widely applied as light-emitter or absorber in many cutting-edge technologies[1,2]. Typically, QDs are synthesized through chemical solution process, and purified with orthogonal solvents, and then processed into thin films or modified with functional molecules for further applications[3]. The surface ligands of QDs play an important role in determining their performance and processing ability due to the well-known ligand-ligand interaction and ligand-solvent interaction[4,5]. In non-polar solvents, ligand-ligand interaction induced steric effects mainly account for dispersion of QDs, while ligand-solvent interaction related electrostatic effects dominate the dispersion of QDs in aqueous solutions[6]. In principle, both steric effects and electrostatic effects contribute to the dispersion of QDs in mixed polar and non-polar solvents. According to the DLVO theory, surface charging mainly determines the electrostatic effects, however, it has been ignored in the previous research, due to the difficulty of the generation of electrical double layer on QDs surface in non-polar solvents[7].

Except for the fundamental research, surface charging of QDs also plays an important role in developing electrophoretic deposition (EPD) processing, in which an applied electric field drives the migration of charged particles in dispersion and deposits them on the electrode to form a coating[8,9]. Considering the EPD application of QDs, it is highly desired to obtain a film with controllable thickness and uniform surface[10–19]. Therefore, it is significant to quantitatively control the surface charge of QDs. According to previous literature, the surface charge of colloidal QDs is closely related to the types and contents of ligands[20–24], solvents[25–28], and external field[29]. Particularly, the EPD results and surface charge states of QDs with the same ligand, as shown in Table S1, are even under controversy. For example, the level of negative surface charge of CdSe based QDs in PGMEA can be controlled by regulating the content of the ligand polyethylene glycol (PEG)-COOH[21]. CdSe nanoplatelets can be separated from the mixture of nanoplatelets and QDs by adding acetone[26]. In addition, carboxylic acid capped CdSe based QDs in toluene can be activated to achieve EPD processing by adding acetonitrile, which was attributed to the ligand removal[25,30]. Up to today, it still lacks a simple and convenient route to gain insight and accurately control the surface charge of QDs in non-polar solvent or mixed solvents.

In this work, we report an alcohol induced surface charging route of colloidal QDs to achieve controllable EPD process. It is found that the zeta potential of OA capped PbSe QDs increases from +1.6 mV to +13.4 mV with the amount of alcohol solvent increasing. By combing the $^1$HNMR, FTIR measurements and the DFT calculations, the positive surface charging of carboxylic acid capped QD is attributed to the electron cloud shift of active hydrogen mediated by intermolecular hydrogen bonds. Considering the influence of surface charging of QDs on their dispersibility, a microscopic mechanism of alcohol-induced EPD of QDs is proposed. Furthermore, we demonstrate the versatility of alcohol induced surface charging strategy, and develop a method for QD separation based on the surface charging effects.

## Results and Discussion

Here the OA capped PbSe QDs were chosen to be the standard material for the investigation. A series of monodisperse PbSe QDs with different sizes were synthesized using the previously reported hot-injection method[31]. Table S2 summarizes the synthesis parameters. Figure S1 and S2 show the transmission electron microscope (TEM) images of PbSe QDs sizing from 5 nm to 22 nm and their corresponding size distributions. According to the literature, the capping ligands of OA, OLA and TOP on QD surface can be characterized using X-ray photoelectron spectroscopy (XPS) and Fourier transform infrared spectroscopy (FTIR). As shown in Figure S3, the signals of N and P elements were not observed in XPS spectra. In addition, stretching vibration absorption peaks corresponding to the COO$^-$ are prominent in FTIR spectra. Based on these results, it is concluded that the as-synthesized PbSe QDs are mainly capped by OA[31].

**EPD of QDs in mixed solvents.** Figure 1a shows the operation procedure of alcohol induced surface charging route and the corresponding EPD process (refer to Experimental Methods for experimental details). It usually contains five steps: i) A certain amount of alcohol solvent was added dropwise to a well-dispersed colloidal QDs solution in toluene or hexane; ii) After ultrasonic for a few minutes, the two-phase solvent gradually become miscible with each other; iii) Colloidal QDs are surface charged for EPD processing; iv) When a fixed voltage was applied on the electrodes, the charged QDs migrate electrophoretically towards the negative electrode; v) As a result, a dense and uniform QDs film was formed (refer to Figure S4 for basic characterization of resulted films). Figure 1b, c show the cross-section SEM images of typical EPD fabricated QDs films. Under conditions induced by different alcohols (ethanol for Figure 1b and methanol for Figure 1c), QDs both pack tightly. And Figure 1d shows the top-surface SEM images of resulted film, where there is smooth surface free of impurities. In addition, oriented films can be obtained using ethanol-induced EPD in the case of balancing dispersion and surface charge. As shown in Figure 1e and S5, there is film with orderly arranged particles, while the QDs film resulted from methanol is always disordered. We further investigate the EPD using other alcohol solvents (methanol, ethanol, isopropanol, n-butanol), similar results can be observed.

The understanding of EPD processing is a prerequisite to improve the quality of QDs film. According to the literature, there are a few techniques investigating the EPD process of QDs, including in-situ optical microscopy[32], equivalent circuit[33], and in-situ mass monitoring[34], where the current-time curves are important tools for qualitatively analysis. Inspired by their works, here we constructed a home-made setup to monitor the deposition flux by combining the in-situ absorption spectra and current-time curves, as shown in Figure S6. By calculating the absorbance under a fixed concentration, the concentration of QDs can be derived from the in-situ absorption spectra. The absorption peaks in the absorbance-time curve correspond to the concentration gradient mediated diffusion of QDs between the electrodes, while the current peaks in the current-time curve correspond to the charge transfer of QDs that have electrophoretically migrated to the electrodes. As shown in Figure 1f, the results indicate that the absorption peaks match the current peaks, with the time difference of peak position less than 5 seconds, suggesting that the occurrence of current peaks in the EPD process are correlated with the deposition flux of QDs.

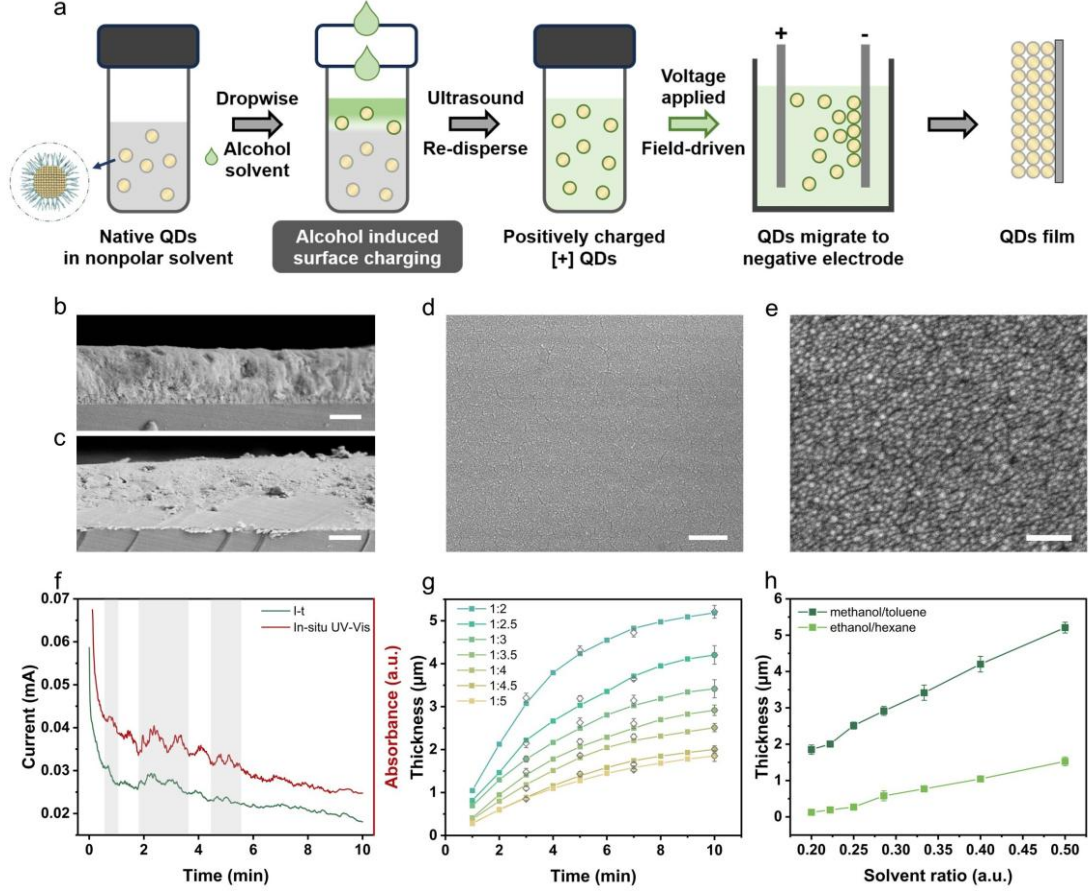

**Figure 1.** Schematic diagrams illustrating the alcohol induced surface charging route and the corresponding EPD results. (a) The surface charging procedure of QDs. (b-c) The cross-section SEM images of PbSe QDs films fabricated by ethanol (scale bar 1μm, top) and methanol (scale bar 2μm, down) induced EPD. (d-e) The top-surface SEM images of smooth PbSe QDs film fabricated by EPD (scale bar 500 nm and 100 nm). (f) The in-situ absorption spectra and current-time curve of a typical EPD process. (g) The thickness-time curves of methanol/toluene induced EPD of PbSe QDs under the electro field strength 40 V/mm, the concentration 0.7 mg/mL, and different solvent ratios. The discrete and hollow data points are experimental verification. (h) The curves of thickness and solvent ratio of PbSe QDs in the solvents methanol/toluene and ethanol/hexane.

To achieve a controllable alcohol-induced surface charging, we first theoretically analyzed the EPD process. According to the EPD model proposed by Hamaker[35], the deposition flux per unit area of the parallel plate electrode can be described by equation 1,

$$J = -D_\infty \frac{dc}{dz} + C\mu_E E_\infty \quad (1)$$

where $J$ is the flux of QD-EPD, $D_\infty$ is the diffusion coefficient of QDs, $C$ is the concentration of QDs, $Z$ is the distance in the direction of the electrode normal, $\mu_E$ is the electrophoretic mobility of QDs, and $E_\infty$ is the strength of applied electric field. Under the condition of significant and stable EPD of QDs, the electrophoretic migration flux $J_e$ is much greater than the diffusion flux $J_c$. In this case, the new equations 2 and 3 can be obtained by defining $\mu_E$ according to Henry equation under Hückel approximation for non-polar systems,

$$J_0 = C\mu_E E_\infty = \frac{2}{3\eta} C\varepsilon\zeta E_\infty \quad (2)$$

$$dw = J_0 dt = \frac{2}{3\eta} C\varepsilon\zeta E_\infty dt \quad (3)$$

where $dw$ is the amount of EPD per unit area and unit time of the charged QDs, $\varepsilon$ is the dielectric constant of the solvent, and $\zeta$ is the zeta potential of the QDs.

Based on the method and Equation 3 above, we investigated the influence of the electric field, QD concentration, deposition time and solvent ratio on the deposition thickness. As shown in Figure S7 (a), the thickness is slightly varied by electric field, while it becomes significant on the premise of ensuring smooth surface when the electric field is higher than 35 V/mm. And as shown in Figure S7 (d) the thickness can be further increased by increasing the concentration of the QDs. The relationship between thickness and deposition time was obtained by integrating the current peak (Figure S8), with the experimental results well verifying it. As shown in Figure 1g, With the extension of the deposition time, the thickness increases linearly within a certain period of time. And then the interfacial resistance of charge transfer increases as the deposits continue to thicken, which not only reduces the current in the deposition circuit, but also causes a certain degree of electric field shielding. As a consequence, the thickness increases into plateau under the combined action of QDs depletion and shielding effect. When the solvent ratio exceeds 1:3, there are large deposition flux and the thickness will reach a plateau in about 10 minutes. Considering the film-forming property, flatness and packing density, the electric field 40v/mm, the concentration 0.7 mg/mL, and deposition time 10 min were chosen as standard parameters.

As summarized in Figure 1h and Table S3, the thickness of resulted QD films can be varied by controlling the volume ratio of alcohol in mixed solvents (polar solvents: methanol, ethanol, isopropanol and n-butanol, non-polar solvents: toluene, hexane and chloroform). Under the standard parameters, the thickness increases with the amount of alcohol solvent. For the colloidal PbSe QDs in methanol/toluene solvents, the thickness of resulted film increases from 1.9 μm to 5.2 μm with the volume ratio of methanol increasing. Figure S9 shows the SEM, AFM, SLCM (scanning laser confocal microscope) images of a typical QDs film fabricated with methanol/toluene solvent. These measurements indicate that the QDs film has smooth surface with the average roughness of 10 nm. Figure S10 shows the SEM, SLCM images of the QD films deposited under different solvent ratios. These images indicates that the QDs films well maintain a smooth surface even with higher alcohol volume ratio. For QDs dispersed in ethanol/hexane, the film thickness increases from 126 nm to 1.5 μm. The deposition thickness of OA capped PbSe QDs increases approximately linearly with the solvent ratio, while the slope decreases as the activity of the hydroxyl hydrogen in the alcohol molecules decreases (refer to Supporting Information for further analysis based on the EPD model). As summarized in Table S4, the EPD fabricated QDs films show smooth and uniform surface, where it is easy to achieve an average roughness of 10nm in the wide thickness ranging from 400 nm to 5 μm (in a 16 μm*16 μm micro-area).

**Mechanism of alcohol-induced positive charging.** To clarify the alcohol induced surface charging of QDs in EPD processing, we investigated the current-time curves of PbSe QDs with different solvents ratio in the mixed solvent of ethanol/hexane. As shown in Figure 2a, the deposition current increases with the amount of alcohol solvent, which can be explained to the reduction of dielectric resistance between the electrodes. Specifically, the current-time curves with increasing ethanol/hexane ratio show distinct peaks during the EPD processing. According to the correlation between absorption peaks and current peaks above, the emerging peaks can be attributed to the enhanced deposition flux. Therefore, we can estimate the total charge transferred between the QDs and electrodes during EPD by integrating the current-time curve (as shown in Figure S11). According to the deposition thickness on a fixed area, we can also estimate the total amount of deposited QDs. Combined with the total charge and total amount, we can further derive the surface charge of a single PbSe QD. As shown in Figure 2b, the surface charge for a single QD increases approximately linear with the solvent ratio of ethanol/hexane. And the proportion of charged QDs induced charge transfer in total charge transfer amount shows two stages, with the solvent ratio 1:3 as inflection point, which can be explained to the sufficient charging of QDs.

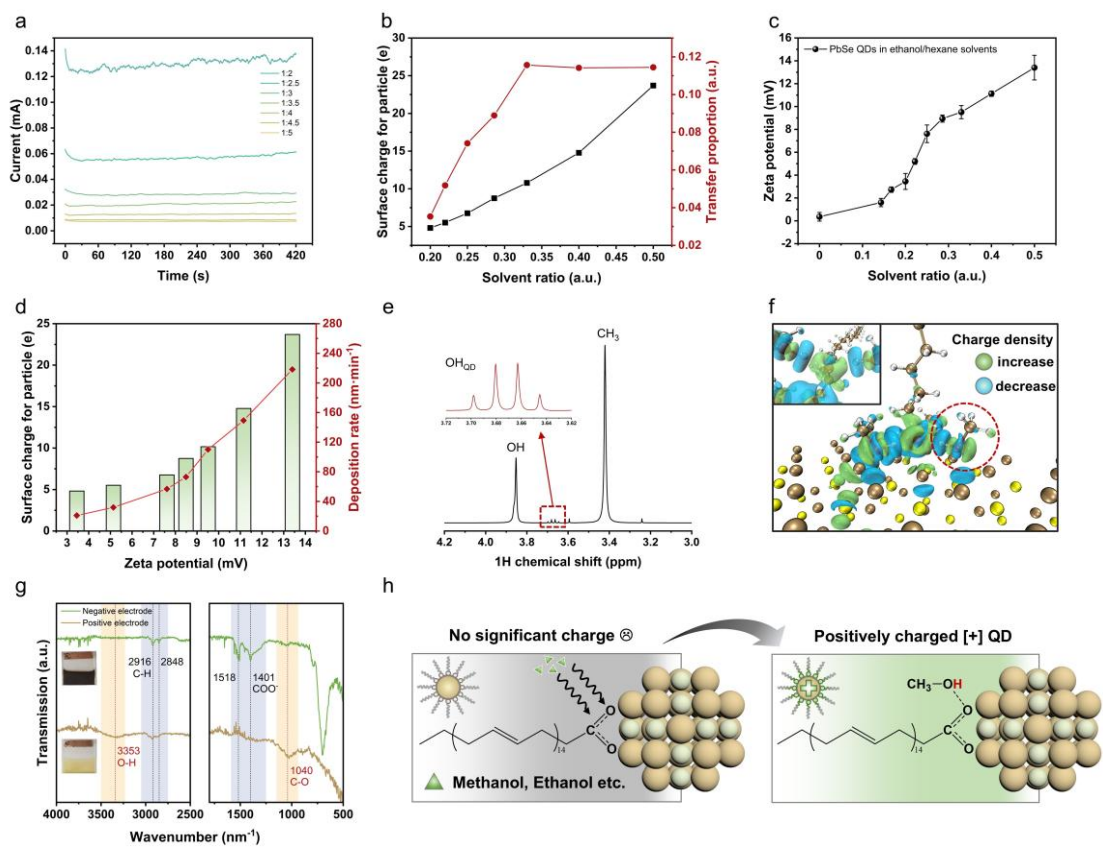

**Figure 2.** Mechanism of alcohol-induced positive charging. (a) The current-time curves of PbSe QDs with different solvents ratio in the ethanol/hexane solvent during EPD. (b) The surface charge of a single QD and the proportion of QDs induced charge transport in the total charge transport as a function of the solvent ratio. (c) The zeta potential of PbSe QDs in ethanol/hexane solvent. (d) Correlation between zeta potential, surface charge and deposition rate of PbSe QDs in ethanol/hexane solvent. (e) The nuclear magnetic spectroscopy of PbSe QDs in methanol/CDCl$_3$ solvent and enlarged view of alcohol interacting with QDs. (f) Schematic diagram of the change in charge density of OA capped PbSe QD surface before and after the addition of alcohol calculated by DFT. Inset: enlarged view of the methanol molecule and carboxylic acid group of OA. (g) FTIR plots of the deposits on positive and negative electrodes. Inset: images of QDs and carbonized deposit on electrodes. (h) Schematic diagram of the interaction between alcohol molecules and OA capped QDs.

Base on the above results, the surface charge of QDs significantly increases with the volume ratio of ethanol increasing. To further understand the influence of alcohol on the surface charge of QDs, we measured the zeta potential of OA capped PbSe QDs in mixed solvents of ethanol/hexane using electrophoretic light scattering. As shown in Figure 2c and S12, the zeta potential of QDs in non-polar solvents such as hexane, toluene, and chloromethane is nearly 0 mV, while the zeta potential of PbSe QDs increased from +1.6 mV to +13.4 mV with the ethanol/hexane solvent ratio increasing from 1:8 to 1:2. Relating the zeta potential to surface charge, as shown in Figure 2d, the results indicate that the higher the zeta potential is, the greater the surface charge and the faster the deposition rate are. This hyphenated analysis provides a new method for measuring the surface charge of colloidal QDs.

To understand the zeta potential change with the alcohol increasing, we characterized the chemical environment of ligands on the surface of QDs using nuclear magnetic resonance hydrogen spectroscopy

($^1$HNMR). As shown in Figure S13, the active hydrogen $^1$HNMR peak of methanol locates at 1.85 ppm. With the increasing of methanol in CDCl$_3$, the peak downshift to 3.36 ppm, resulting in a broad peak when the amount increases up to volume fraction 20%. The downshift of the active hydrogen can be attributed to the reduced shielding effect originating from the enhanced hydrogen bonding effect between alcohol molecules. In the presence of QDs, the peak of the active hydrogen isolates and downshift to 3.85 ppm. Intermolecular hydrogen bonding occurs between carboxylic acid ligands and methanol, and the exchange of active hydrogen is limited and slowed down. As a result, as shown in Figure 2f, the alcohol hydroxyl group interacting with the ligand exhibits coupled splitting of the methyl group at 3.67 ppm. Since the intermolecular hydrogen bonding between QDs and methanol is weaker than that between methanol molecules, the shielding effect of active hydrogen is enhanced, and the splitting NMR peak moves to the high field.

The influence of methanol on OA capped PbSe QDs were further studied by applying first-principles calculation. Figure 2h show the charge density difference of the OA capped PbSe QDs before and after the presence of methanol at 300 K. The charge density of the proton hydrogen decreases significantly (green regions), while the charge density of the carbonyl oxygen on the carboxyl group increases significantly (blue regions). These results indicate that the active hydrogen of the alcohol is closely correlated and even coupled to the carboxyl group, due to the intermolecular hydrogen bonding.

To further illustrate the influence of intermolecular hydrogen bonding between the methanol and QDs on EPD process, we characterized the deposits on the electrodes using FTIR and XRD measurements. As shown in insets of Figure 2g, the black negative electrode is covered by dense QDs film, while the positive electrode is covered by unknown yellow deposit. As shown in Figure 2g, the FTIR spectra and XRD patterns of the black deposits confirmed the EPD of QDs on the negative electrode. Based on the FTIR spectra in Figure 2g and S14, yellow deposit on positive electrode exhibits significant C-O stretching vibration absorption peak around 1040 cm$^{-1}$ and O-H stretching vibration absorption near 3353 cm$^{-1}$. The XRD measurements of the yellow deposit show characteristic diffraction peaks at 44 degrees (see Figure S15). According to these results, the yellow deposit can be assigned to the carbonized side products. Based on the above results, as shown in Figure 2h, it can be deduced that the surface charging of OA capped PbSe QDs originates from the intermolecular hydrogen bonding between the carboxylic group and alcohol, which resulting in a shift of the active hydrogen electron cloud.

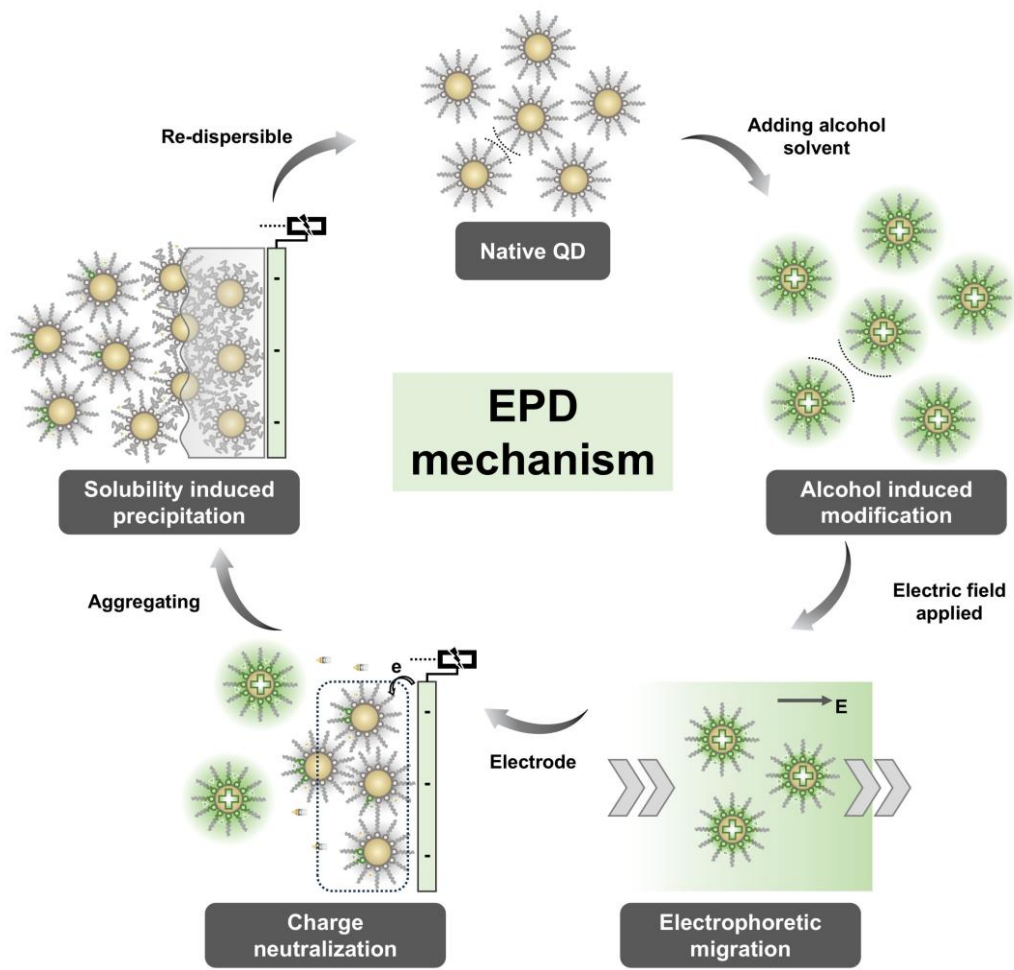

**Figure 3.** Schematic diagrams illustrating the mechanism of alcohol-induced EPD of QDs.

**Microscopic mechanism of alcohol-induced EPD.** Based on the above characterization and previous literature, the mechanism of alcohol-induced EPD of QDs was further investigated. As shown in Figure 3, the EPD process can be divided into five stages, including i) pre-dispersion, ii) surface charging, iii) electrophoretic migration, iv) charge neutralization and v) precipitation. It is noted that the surface charging of QDs significantly affect both the electrophoretic migration and deposition process. In the stage ii), the surface charging of QDs affect their dispersibility due to the synergistic effects of steric and electrostatic interactions. As shown in Figure S16 and Table S5, the scattering light in the absorption spectrum decreased. Meanwhile the dynamic light scattering (DLS) measurements indicate that the polydispersity index decreased significantly from 0.112 to 0.047, with the adding of 5 % methanol in volume. The synergistic effect of steric effect and electrostatic effect increases spacing between QDs with minimal impact on the polarity of solvent. In the stage iii), the surface charging affect the migration mobility and the deposition thickness of QDs. Under the applied electric field, charged QDs undergo directional migration electrophoretically, during which the methanol molecule is partially dissociated, and the active hydrogen is highly coupled with the carboxyl group of OA. In the stage iv), the positively charged QDs undergo charge transfer on the negative electrode and transform into neutral. The methanol molecule completely dissociates. In the stage v), the solubility of QDs in methanol/toluene solvents decreases after they converted to neutral, resulting in a charge neutralization mediated aggregation. The aggerates precipitate onto the electrode to form QDs film.

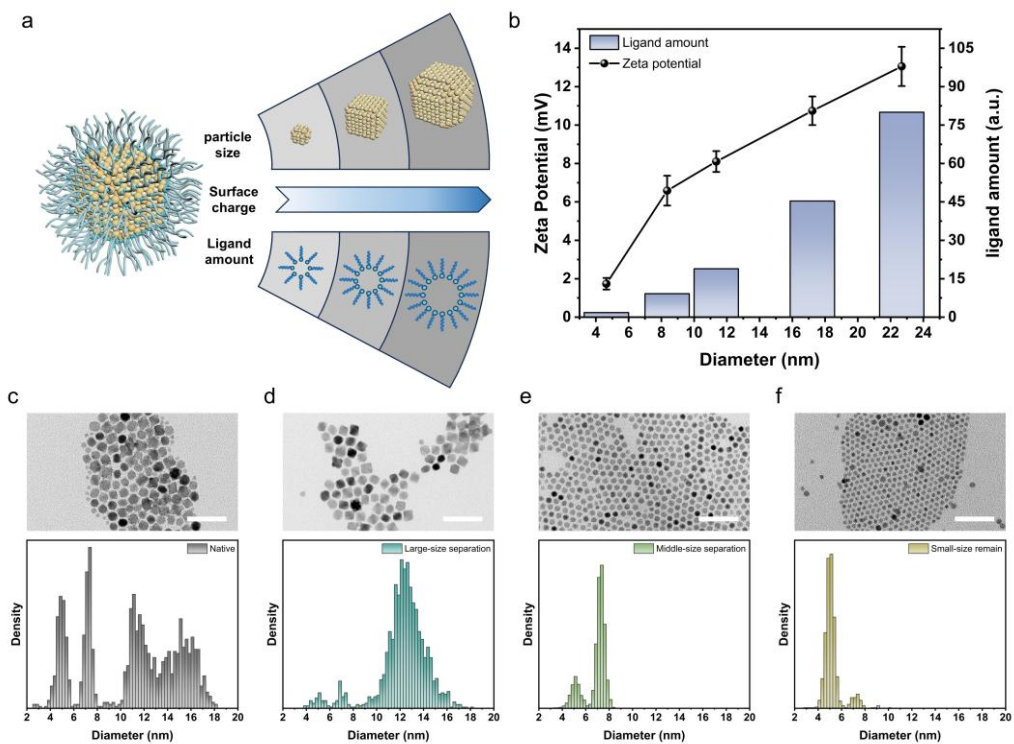

**Figure 4.** Size dependence of surface charging properties. (a) Schematic diagram illustrating that surface charging originates from the ligand amount. (b) The curve of zeta potential and ligand amount of a single QD as a function of size. TEM image and the distribution diagram of (c) pre-dispersed PbSe QDs mixture, and (d-f) the separated three monodisperse samples. The scale bar is 50 nm. Each set of data is a statistical result of approximately 10,000 QDs obtained by a detection-segmentation model and its accompanying software[36].

**A size-selective separation protocol.** It has been confirmed that the surface charging properties of QDs is derived from the interaction between active hydrogen and OA ligands. Consequently, the surface charging properties of QDs is expected to exhibit size dependence, because the ligand amount, in other words the proportion of surface atoms, are size related. The size, ligand amount, and surface charge of QDs exhibit logical relationships. As shown in Figure 4a, the ligand amount of a single QD increases with the size, and the alcohol induced surface charge increases with the ligand amount, which shows a size-ligand-charge dependence. We measured the zeta potential of OA capped PbSe QDs of different sizes under the solvent ratio ethanol/hexane-1:4. As summarized Figure 4b, the zeta potentials of PbSe QDs are 1.74 mV, 6.59 mV, 8.10 mV, 10.74 mV, 13.05 mV for the samples with average diameter of 4.64 nm, 8.35 nm, 11.35 nm, 17.22 nm, 22.68 nm respectively. Except for the smallest PbSe QD with a diameter of 4.64 nm, the zeta potential of alcohol induced PbSe QDs is almost linearly related to size. We further calculated the ligand amount of PbSe QDs by applying thermal gravimetric analysis (TGA)[31]. As shown in Figure 4b and Table S6, the calculated ligand amount on a single QD increases with the size increasing, which is correlated with the size dependence of alcohol-induced zeta potential.

Building upon the understanding into microscopic mechanism of alcohol-induced EPD, we developed a size-selective separation protocol. Three PbSe QDs with average diameters (4.9 nm, 7.3 nm, 13.6 nm) were mixed as a target sample for size-selective separation. Figure 4c showed the typical TEM image of the target sample. By designing a protocol with the electric field, solvent ratio and deposition time, we can separate QDs with different diameters. The details are described in Experimental section and Figure S17. As shown in Figure 4d, e, f, the target mixed QDs can be subsequently separated into three monodisperse samples with the average diameters 5.0 nm, 7.3 nm, 12.7 nm.

To demonstrate the universality of the alcohol-induced surface charging properties of colloidal QDs, a series of colloidal QDs with different ligands were synthesized (refer to Supporting Information for synthesis methods). As shown in Figure S18, the zeta potential of these colloidal QDs were measured under the same solvent ratio ethanol/hexane-1:4. The results indicate that the surface charge properties are universal to different types of OA capped QDs such as PbS, $Cu_2Se$, $CuInSe_2$, and $CsPbBr_3$, with a considerably zeta potential value compared to the polar dispersion systems. Additionally, it is also applicable to QDs capped with other carboxylic acid ligands (2-HDA, OCTA), and core-shell QDs which are mainly capped with OA, containing TOP, DPP and other mixed ligand components.

## Conclusion

In summary, we reported the alcohol induced surface charging route and the corresponding EPD processing of QDs. By regulating the deposition time, solvent ratio, concentration, and electric field, a controllable EPD process is achieved, with the deposition thickness of PbSe QDs in methanol/toluene solvents ranging from 400 nm to 5 μm simply. These results illustrate the surface positive charging of QDs. The surface charging of PbSe QDs was investigated by zeta potential, $^1$HNMR, FTIR, and DFT calculation. It was found that the zeta potential of OA capped PbSe QDs increases from +1.6 mV to +13.4 mV with the amount of alcohol solvent increasing. The alcohol-induced surface positive charging can be explained to the electron cloud shift of active hydrogen mediated by intermolecular hydrogen bonds. Furthermore, we proposed the microscopic mechanism to explain the influence of surface charging on alcohol-induced EPD of QDs, which include five stages of pre-dispersion, surface charging, electrophoretic migration, charge neutralization and precipitation. Based on the size dependence of alcohol-induced zeta potential, a size-selective separation protocol was developed. A target sample of mixed PbSe QDs with three average diameter 5.1 nm, 7.8 nm and 12.6 nm were separated into three monodisperse samples by electrophoretic modulation from the electric field, solvent ratio and deposition time. In addition, we also demonstrate the versatility of alcohol-induced surface charging properties of different types of QDs and QDs with mixed ligands.

## Experimental Methods

**Surface charging of QDs.** The alcohol-induced surface charging process of QDs was achieved in two steps. First, as-synthetic QDs were dispersed in non-polar solvents such as hexane and toluene. Alcohol solvents are added dropwise, followed by sonicating and slowly shaking until there are no longer phase-splitting, where the alcohol solvents and CQDs are completely miscible in each other.

**Fabrication of QD films by EPD.** For a typical EPD process, positively charged PbSe QDs (0.3-2.0 mg mL$^{-1}$) dispersed in methanol/toluene-1:4 in volume were injected into a home-made deposition tank. Two parallel electrodes (ITO substrates generally, silicon substrates can be used for better orientation) were vertically fixed in the solution with a spacing of 5 mm. After applying a bias of 50-200 V for 10 min, a black PbSe QD film was formed on the negative electrode, while a yellow carbonized deposit was formed on the positive electrode. The deposited QDs film could be removed from the solution and redissolved in toluene.

**Size-selective separation of PbSe QDs**. The separations of PbSe QDs were achieved through a controlled EPD process. Initially, an electric field of 10 V/mm and a solvent ratio of methanol/toluene-1:6 was used to separate 12.7 nm large-sized QDs. Subsequently, an electric field of 20 V/mm and a solvent ratio of methanol/toluene-1:4 was applied for the separation of 7.3 nm medium-sized QDs. Finally, 5.0 nm small-sized QDs were left after an electric field of 40 V/mm and a solvent ratio of methanol/toluene-1:2. The detailed separation results are summarized in Figure S17.

**References**

1. García De Arquer, F. P.; Talapin, D. V.; Klimov, V. I.; Arakawa, Y.; Bayer, M.; Sargent, E. H. Semiconductor Quantum Dots: Technological Progress and Future Challenges. *Science*. **2021**, *373*, eaaz8541.
2. Efros, A. L.; Brus, L. E. Nanocrystal Quantum Dots: From Discovery to Modern Development. *ACS Nano*. **2021**, *15*, 6192–6210.
3. Kagan, C. R.; Lifshitz, E.; Sargent, E. H.; Talapin, D. V. Building Devices from Colloidal Quantum Dots. *Science*. **2016**, *353*, aac5523.
4. Scholes, G. D. Controlling the Optical Properties of Inorganic Nanoparticles. *Adv. Funct. Mater.* **2008**, *18*, 1157–1172.
5. Calvin, J. J.; Brewer, A. S.; Alivisatos, A. P. The Role of Organic Ligand Shell Structures in Colloidal Nanocrystal Synthesis. *Nat. Synth.* **2022**, *1*, 127–137.
6. Boles, M. A.; Ling, D.; Hyeon, T.; Talapin, D. V. The Surface Science of Nanocrystals. *Nat. Mater.* **2016**, *15*, 141–153.
7. Dickerson, J. H., Boccaccini, A. R., Electrophoretic Deposition of Nanomaterials; Springer New York: New York, NY, 2012.
8. Van Der Biest, O. O.; Vandeperre, L. J. ELECTROPHORETIC DEPOSITION OF MATERIALS. *Annu. Rev. Mater. Sci.* **1999**, *29*, 327–352.
9. Besra, L.; Liu, M. A Review on Fundamentals and Applications of Electrophoretic Deposition (EPD). *Prog. Mater Sci.* **2007**, *52*, 1–61.
10. Islam, M. A.; Xia, Y.; Steigerwald, M. L.; Yin, M.; Liu, Z.; O'Brien, S.; Levicky, R.; Herman, I. P. Addition, Suppression, and Inhibition in the Electrophoretic Deposition of Nanocrystal Mixture Films for CdSe Nanocrystals with γ-$Fe_2O_3$ and Au Nanocrystals. *Nano Lett.* **2003**, *3*, 1603–1606.
11. Islam, M. A.; Xia, Y.; Telesca, D. A.; Steigerwald, M. L.; Herman, I. P. Controlled Electrophoretic Deposition of Smooth and Robust Films of CdSe Nanocrystals. *Chem. Mater.* **2004**, *16*, 49–54.
12. Whan Lee, S.; Zhang, D.; Herman, I. P. Rapid and Multi-Step, Patterned Electrophoretic Deposition of Nanocrystals Using Electrodes Covered with Dielectric Barriers. *Appl. Phys. Lett.* **2014**, *104*, 053113.
13. Otelaja, O. O.; Ha, D.-H.; Ly, T.; Zhang, H.; Robinson, R. D. Highly Conductive $Cu_{2-x}S$ Nanoparticle Films through Room-Temperature Processing and an Order of Magnitude Enhancement of Conductivity via Electrophoretic Deposition. *ACS Appl. Mater. Interfaces*. **2014**, *6*, 18911–18920.
14. Yu, Y.; Yu, D.; Orme, C. A. Reversible, Tunable, Electric-Field Driven Assembly of Silver Nanocrystal Superlattices. *Nano Lett.* **2017**, *17*, 3862–3869.
15. Ravi, V. K.; Scheidt, R. A.; DuBose, J.; Kamat, P. V. Hierarchical Arrays of Cesium Lead Halide Perovskite Nanocrystals through Electrophoretic Deposition. *J. Am. Chem. Soc.* **2018**, *140*, 8887–8894.
16. Aniskevich, Y.; Radchanka, A.; Antanovich, A.; Prudnikau, A.; Quick, M. T.; Achtstein, A.

W.; Jo, J. H.; Ragoisha, G.; Artemyev, M.; Streltsov, E. Electrophoretically-Deposited CdSe Quantum Dot Films for Electrochromic Displays and Smart Windows. *ACS Appl. Nano Mater.* **2021**, *4*, 6974–6984.

17. Zhang, Y.; Pham, X.-M.; Keating, T.; Jia, N.; Mullen, A.; Laishram, D.; Gao, M.-Y.; Corbett, B.; Liu, P.; Sun, X. W.; Soulimane, T.; Silien, C.; Ryan, K. M.; Ma, Z.; Liu, N. Highly Efficient Inverted Light-Emitting Diodes Based on Vertically Aligned CdSe/CdS Nanorod Layers Fabricated by Electrophoretic Deposition. *ACS Appl. Mater. Interfaces.* **2024**, *16*, 10459–10467.

18. Liu, G.; Wu, X.; Xiong, F.; Yang, J.; Liu, Y.; Liu, J.; Li, Z.; Qin, Z.; Deng, S.; Yang, B.-R. Fluorescent, Multifunctional Anti-Counterfeiting, Fast Response Electrophoretic Display Based on TiO2/CsPbBr3 Composite Particles. *Light Sci. Appl.* **2024**, *13*, 198.

19. Li, X.; Zhao, J.; Xiao, H.; Zhang, H.; Zhou, M.; Zhang, X.; Yan, X.; Tang, A.; Chen, L. Multiparticle Synergistic Electrophoretic Deposition Strategy for High-Efficiency and High-Resolution Displays. *ACS Nano.* **2024**, *18*, 17715–17724.

20. Jia, S.; Banerjee, S.; Herman, I. P. Mechanism of the Electrophoretic Deposition of CdSe Nanocrystal Films: Influence of the Nanocrystal Surface and Charge. *J. Phys. Chem. C.* **2008**, *112*, 162–171.

21. Zhao, J.; Chen, L.; Li, D.; Shi, Z.; Liu, P.; Yao, Z.; Yang, H.; Zou, T.; Zhao, B.; Zhang, X.; Zhou, H.; Yang, Y.; Cao, W.; Yan, X.; Zhang, S.; Sun, X. W. Large-Area Patterning of Full-Color Quantum Dot Arrays beyond 1000 Pixels per Inch by Selective Electrophoretic Deposition. *Nat. Commun.* **2021**, *12*, 4603.

22. Park, Y.; Kang, H.; Jeong, W.; Son, H.; Ha, D.-H. Electrophoretic Deposition of Aged and Charge Controlled Colloidal Copper Sulfide Nanoparticles. *Nanomaterials.* **2021**, *11*, 133.

23. Song, M.; Kim, Y.; Baek, D. S.; Kim, H. Y.; Gu, D. H.; Li, H.; Cunning, B. V.; Yang, S. E.; Heo, S. H.; Lee, S.; Kim, M.; Lim, J. S.; Jeong, H. Y.; Yoo, J.-W.; Joo, S. H.; Ruoff, R. S.; Kim, J. Y.; Son, J. S. 3D Microprinting of Inorganic Porous Materials by Chemical Linking-Induced Solidification of Nanocrystals. *Nat. Commun.* **2023**, *14*, 8460.

24. Xiao, H.; Zhao, J.; Li, X.; Zhang, H.; Zhou, M.; Cao, W.; Yan, X.; Zhang, X.; Sun, X. W.; Chen, L. "Nanoscale Electric Vehicle" for the Patterning of Nanomaterials: Selective Electrophoretic Deposition of Programmable Silica Composite Nanoparticles. *Nano Energy.* **2024**, *128*, 109906.

25. Song, K. W.; Costi, R.; Bulović, V. Electrophoretic Deposition of CdSe/ZnS Quantum Dots for Light-Emitting Devices. *Adv. Mater.* **2013**, *25*, 1420–1423.

26. Lhuillier, E.; Hease, P.; Ithurria, S.; Dubertret, B. Selective Electrophoretic Deposition of CdSe Nanoplatelets. *Chem. Mater.* **2014**, *26*, 4514–4520.

27. Kang, H.; Park, Y.; Hong, Y.-K.; Yoon, S.; Lee, M.-H.; Ha, D.-H. Solvent-Induced Charge Formation and Electrophoretic Deposition of Colloidal Iron Oxide Nanoparticles. *Surf. Interfaces.* **2021**, *22*, 100815.

28. Xu, X.; Kweon, K. E.; Keuleyan, S.; Sawvel, A.; Cho, E. J.; Orme, C. Rapid In Situ Ligand-Exchange Process Used to Prepare 3D PbSe Nanocrystal Superlattice Infrared Photodetectors. *Small.* **2021**, *17*, 2101166.

29. Luo, C.; Ding, Y.; Ren, Z.; Wu, C.; Huo, Y.; Zhou, X.; Zheng, Z.; Wang, X.; Chen, Y. Ultrahigh-Resolution, High-Fidelity Quantum Dot Pixels Patterned by Dielectric Electrophoretic Deposition. *Light Sci. Appl.* **2024**, *13*, 273.

30. Hassinen, A.; Moreels, I.; De Nolf, K.; Smet, P. F.; Martins, J. C.; Hens, Z. Short-Chain Alcohols Strip X-Type Ligands and Quench the Luminescence of PbSe and CdSe Quantum Dots, Acetonitrile Does Not. *J. Am. Chem. Soc.* **2012**, *134*, 20705–20712.
31. Gu, K.; Wu, H.; Su, J.; Sun, P.; Tan, P.-H.; Zhong, H. Size Dependent Specific Heat Capacity of PbSe Nanocrystals. *Nano Lett.* **2024**, *24*, 4038–4043.
32. Kornhuber, K.; Kavalakkatt, J.; Lin, X.; Ennaoui, A.; Lux-Steiner, M. Ch. In Situ Monitoring of Electrophoretic Deposition of Cu2ZnSnS4 Nanocrystals. *RSC Adv.* **2013**, *3*, 5845–5850.
33. Dillon, A. D.; Mengel, S.; Fafarman, A. T. Influence of Compact, Inorganic Surface Ligands on the Electrophoretic Deposition of Semiconductor Nanocrystals at Low Voltage. *Langmuir*. **2018**, *34*, 9598–9605.
34. Panta, K. R.; Orme, C. A.; Flanders, B. N. Quantitatively Controlled Electrophoretic Deposition of Nanocrystal Films from Non-Aqueous Suspensions. *J. Colloid Interface Sci.* **2023**, *636*, 363–377.
35. Hamaker, H. C. Formation of a Deposit by Electrophoresis. *Trans. Faraday Soc.* **1940**, *35*, 279–287.
36. Gu, K.; Liang, Y.; Su, J.; Sun, P.; Peng, J.; Miao, N.; Sun, Z.; Fu, Y.; Zhong, H.; Zhang, J. Deep Learning Models for Colloidal Nanocrystal Synthesis. *arXiv e-prints*. **2024**, arXiv:2412.10838.

## Acknowledgement

This work was supported by the National Natural Science Foundation of China (U23A20683, H.Z.), the Natural Science Foundation of Beijing Municipality (No. Z210018), Beijing Municipal Science & Technology Commission, Administrative Commission of Zhongguancun Science under Park No. Z231100006023018, and the National Natural Science Foundation of China (62422501).

We thank Mr. Jiayu Yu for his valuable advice on surface charging characterization, and Mr. Mingrui Liu, Ms. Min Yang, Mr. Tao Wang for their support in synthesis.

## Supplementary Information
**Supplementary Methods**

**Materials**. All commercially available chemicals were used without further purification. Lead oxide (PbO, 99.9%, Aladdin) copper(I) chloride (CuCl, 98%, Energy Chemical), copper(I) iodide (CuI, 98%, Energy Chemical), indium (III) acetate (In(Ac)3, 99.99%, Macklin), zinc(II) acetate (Zn(Ac)2, 99.5%, Macklin), lead(II) bromide (PbBr2, 99.9%, Macklin), caesium carbonate (Cs2CO3, 99.9%, Macklin), selenium powder (Se, 99.99%, Aladdin), sulfur powder (S, 99.99%, Aladdin), oleic acid (OA, 90%, Energy Chemical), trioctylphosphine (TOP, 90%, Energy Chemical), 1-octadecene (ODE, 90%, Alfa Aesar), oleylamine (OLA, 90%, Energy Chemical), octanoic acid (OCTA, 99%, Energy Chemical), 1-dodecanethiol (DDT, 98%, Aladdin) hexane (99%, Sigma-Aldrich), toluene (99%, Sigma-Aldrich), tetrachloroethylene (TCE, 99.9%, Aladdin), dichloromethane (DCM, 99.9%, Aladdin), N,N-Dimethylformamide (DMF, ≥99.9%, Aladdin), methanol (99.5%, Macklin), ethanol (99.9%, Sigma-Aldrich), isopropanol (99.9%, Sigma-Aldrich), n-butanol (99.9%, Sigma-Aldrich), acetonitrile (99.9%, Sigma-Aldrich) acetone (99.9%, Sigma-Aldrich) and acetic acid (99.7%, Macklin) were used in quantum dots synthesis, purification and solvent induced modification. The synthesis method of mainly OA capped PbSe, PbS, Cu2Se, CuInSe2, CsPbBr3 QDs and multiple ligands capped CdZnSeS, ZnSe based core-shell QDs are mentioned in Supplementary methods. The mainly 2-hexyldecanoic acid (HDA) capped CdSe based green QDs were purchased from Beida Jubang Technology Co., Ltd. The polyethylene glycol (PEG) modified CdSe based QDs and Phenethylammonium bromide (PEABr) were purchased from ZhiJing Technology Co., Ltd.

**Synthesis of PbSe and PbS quantum dots**. TOPSe (selenium powder dissolved in TOP) were used as the precursors by heating the TOP and Se powder at 40 °C for 30 min. Taking the synthesis of OA capped PbSe quantum dots with a diameter of 8.35 nm as an example, 1.116 g of PbO, 7.05 mL of OA, 1.65 mL of OLA and 15 mL of ODE were degassed under vacuum at 100 °C for an hour until a light yellow clarification solution was achieved. Then the solution was heated to 150 °C under flowing nitrogen, at which temperature 2.5 mL of a 2M solution of TOPSe was rapidly injected to trigger the nucleation. After 15 min, the reaction was terminated by a water bath. The obtained quantum dots were purified by two rounds of precipitation-redispersion with acetone and methanol (1:1)/hexane, dried completely and stored in nitrogen. The synthesis parameters for quantum dots with different sizes are shown in Table S2. Among them, caprylic acid capped PbSe quantum dots were dispersed in DCM. The synthesis process of PbS quantum dots in consistent with that of PbSe quantum dots, except that different synthesis parameters are used.

**Synthesis of $Cu_2Se$ quantum dots**. Cu2Se quantum dots were synthesized using an adapted hot-injection method referred to previously report[1]. 1 mmol Se powder was mixed with 5 mL OA under vacuum at 100 °C for 10 min, and the mixture was heated to 300 °C until the solution became clear. Then the solution was stored at 140 °C as Se precursor. 1 mmol CuCl, 5 mL OA and 5 mL OLA were loaded in a three-necked flask under flowing nitrogen and heated to 220 °C until the solution turned to yellow and clear. 5 mL Se precursor was then rapidly injected and heated to 270 °C. After 10 min at 270 °C, the mixture was cooled to room temperature naturally. The obtained quantum dots were purified by two rounds of precipitation-redispersion with ethanol and chloroform, and finally dispersed in TCE.

**Synthesis of $CuInSe_2$ quantum dots.** CuInSe$_2$ quantum dots were synthesized using an adapted hot-injection method referred to previously report[2]. 0.4 mmol Se powder was dissolved in the mixture of 0.7 mL OLA and 0.3 mL DDT as the Se precursor. 0.18 mmol CuI, 0.2 mmol In(Ac)$_3$, 1 mL DDT and 10 mL ODE were loaded in a 25-mL three-necked flask and dried under vacuum at 120 °C for 30 min to prepare the Cu precursor. Then 0.5 mL OLA and 0.75 mL OA were injected to the flask under flowing nitrogen. After heating for 30 min, the Cu precursor was heated to 220 °C, and 1 mL Se precursor was rapidly injected to trigger the nucleation. After 20 min, the reaction was terminated by a water bath. The obtained quantum dots were purified by two rounds of precipitation-redispersion with ethanol (redispersed in hexane), dried completely and stored in nitrogen.

**Synthesis of $CsPbBr_3$ quantum dots.** 0.069 g of PbBr$_2$ and 5 mL of ODE were loaded into a 25-mL three-necked flask and dried under vacuum at 120 °C for 1 h. Separately, to prepare the Cs-oleate precursor, 0.814 g of Cs$_2$CO$_3$, 2.5 mL of OA, and 40 mL of ODE were loaded into a 100-mL three-necked flask and dried at 120 °C for 1 h, and then heated under nitrogen to 150 °C until all Cs$_2$CO$_3$ reacted with OA. The Cs-oleate precursor was preheated above 100 °C before injection. Once the PbBr$_2$ flask was under vacuum for 1 h, 0.5 mL of OLA and 0.5 mL of OA were injected at 120 °C under nitrogen. After complete dissolution of PbBr$_2$, the temperature was raised to 180 °C and 0.4 mL of Cs-oleate solution was quickly injected. After 5 s, the reaction mixture was cooled by the ice-water bath. Then, the solution was cleaned by centrifugation at 10,000 rpm for 20 min twice. The supernatant was removed, and the precipitates were dispersed in toluene.

**Synthesis of CdZnSeS based quantum dots.** Cd$_x$Zn$_{1-x}$Se$_y$S$_{1-y}$/ZnS quantum dots were synthesized using an adapted hot-injection method referred to previously report[3]. 1.3 mmol Se and 13.3 mmol S powders were dissolved in 10 mL TOP as the Se/S precursor. 15 mmol Zn(Ac)$_2$, 15 mL DDT, 10 mL OLA and 15 mL ODE were loaded in a three-necked flask and heated at 80 °C for 20 min until dissolved, followed by keeping warm at 120 °C under flowing nitrogen as the shell precursor. 0.12 mmol CdO, 1.2 mmol Zn(Ac)$_2$, 1.8 mL OA and 6 mL ODE were loaded in a three-necked flask and heated to 120 °C for 20 min under vacuum. Then it was heated to 300 °C under flowing nitrogen, followed by rapidly injection of 1 mL Se/S precursor to synthesis the Cd$_x$Zn$_{1-x}$Se$_y$S$_{1-y}$ core. After 10 min, the reaction solution was heated to 280 °C, and 4 mL shell precursor was dropwise injected at a rate of 0.1 mL·min$^{-1}$ using a syringe pump. After the reaction was completed, obtained Cd$_x$Zn$_{1-x}$Se$_y$S$_{1-y}$/ZnS quantum dots were naturally cooled to 60 °C and purified by three rounds of precipitation-redispersion with acetone and methanol 7:3, acetone and methanol 3:7, methanol and acetic acid 1:4 in volume (redispersed in toluene). Quantum dots were dried at 60 °C under vacuum to get powder.

**Synthesis of ZnSe based quantum dots.** ZnCdSe/ZnCdSeS/ZnS quantum dots were synthesized using the previously reported hot-injection method Q2[4].

**Synthesis of Br⁻ capped PbSe quantum dots.** Br- capped PbSe quantum dots were obtained using the phase-transfer based ligand exchange method. 1 mmol PEABr was dissolved in 9 mL DMF, and 1.5 mg/mL OA capped PbSe quantum dots were dispersed in 6 mL hexane. The colloidal PbSe quantum dots were added to PEABr solution and it is phase separated. The two-phase mixture was stirred at 200 rpm at 50 °C for 5 min under nitrogen, and the quantum dots were gradually transferred into the DMF solution. Discard the upper hexane, and the quantum dots in DMF were washed by two rounds of 3 mL hexane, stirring at 500 rpm for 2 min. The obtained Br⁻ capped PbSe quantum dots with PEA$^+$ counter-ion diffusion layer were used for measurements and electrophoretic deposition without further precipitation-redispersion.

**Synthesis of ZnMgO based nanoparticles.** ZnMgO nanoparticles were synthesized and later modified by siloxane using the previously reported method[5].

**Simulations details.** Simulations are performed within the CP2K program using the Quickstep module. The perdew-Burke-Ernzerhof (PBE) functional was used for the exchange-correlation potential. The interaction between the ionic cores and valence electrons was described using Goedecker-Teter-Hutter (GTH) pseudopotentials. The valence electron wavefunctions were expanded using a Double-Zeta-Plus-Polarization (DZVP) Gaussian basis set. The plane-wave cutoff energy was set to 400 Ry to ensure convergence of the total energy. Structural optimization was carried out using the BFGS algorithm, with convergence criteria of force change less than 0.01 eV·Å$^{-1}$.

**Supplementary Figures and Tables**

**Supplementary Table 1.** Literature survey of QD-EPD in non-polar dispersion and polar dispersion, revealing the surface charging properties of QDs.

| System Composition Ligand Size | Solvent | EPD result | Surface charging properties | Citation |
|---|---|---|---|---|
| CdSe-TOPO | Hexane | Both | + & - | [6] Herman 2002 |
| $Fe_2O_3$-OA | Hexane | Both | + & - | [7] Herman 2003 |
| Au-SH | Hexane | None | Unclear | |
| $Eu_2O_3$-TOPO | Hexane | Both | + & - | [8] Dickerson 2006 |
| CdS-ODPA+TOPO 8*100nm rod | Toluene | Positive electrode | - | [9] Ryan 2009 |
| CdSe-TOPO+TBP | Toluene | Both | + & - | [10] Zaban 2010 |
| PbSeS-OA PbS-OA | Toluene | Both | + & - | [11] Benehkohal 2012 |
| CdSe/ZnS-OA | Chloroform/acetonitrile | Both | + & - | [12] Bulović 2013 |
| CdSe-OA | Hexane/acetone | Positive electrode | - | [13] Dubertret 2013 |
| $Cu_2S$-OLA | Hexane | Positive electrode | - | [14] Robinson 2014 |
| $CuInSe_2$-OLA | Toluene/ethanol | Positive electrode | - | [15] Fafarman 2016 |
| Ag-SH | Toluene | Negative electrode | + | [16] Orme 2017 |
| $CsPbX_3$-$PbSO_4$/OA | Hexane | Positive electrode | - | [17] Kamat 2018 |
| $CsPbBr_3$-OA/PVP | Toluene | Positive electrode | - | [18] Im 2022 |
| PbSe-OA | Toluene/acetonitrile | Negative electrode | + | [19] Orme 2021 |
| **PbSe-OA** | **Toluene/acetonitrile** | **Positive electrode** | **-** | **This work** |
| **PbSe-OA** | **Toluene/ethonal** | **Negative electrode** | **+** | |
| $CuInSeS$-$Se^{2-}$ | DMF | Positive electrode | - | [20] Fafarman 2018 |
| PbSe-$Br^-$ | DMF | Positive electrode | - | This work |
| PbSe-PEABr | DMF | Negative electrode | + | |
| CdSe-SPEGCOOH | PGMEA | Positive electrode | - | [21] Sun 2021 |

**Supplementary Table 2.** Synthesis conditions and size of PbSe and other different kinds of quantum dots.

| Samples | | Amount of precursor and ligand (mmol) | Reaction temperature (°C) | Reaction time (min) | Average size (nm) and size distribution (%) |
|---|---|---|---|---|---|
| PbSe | S1 | PbO(5), Se(5), ODE(46.88), OA(22.21), OLA(5.015), TOP(5.605) | 140 | 5 | 4.64(8.30) |
| PbSe | S2 | PbO(5), Se(5), ODE(46.88), OA(22.21), OLA(5.015), TOP(5.605) | 150 | 15 | 8.35(5.69) |
| PbSe | S3 | PbO(5), Se(5), ODE(120.2), OA(22.21), OLA(5.015), TOP(5.605) | 160 | 40 | 11.35(6.42) |
| PbSe | S4 | PbO(5), Se(5), ODE(120.2), OA(22.21), OLA(5.015), TOP(5.605) | 240 | 10 | 17.22(15.81) |
| PbSe | S5 | PbO(5), Se(5), ODE(120.2), OA(22.21), OLA(5.015), TOP(5.605) | 240 | 60 | 22.68(6.68) |
| PbSe | | PbO(5), Se(5), ODE(120.2), OCTA(20.45), OLA(5.015), TOP(5.605) | 180 | 40 | 16.62(19.23) |
| PbS | | PbO(4), S(1), ODE(62.5), OA(17.8), OLA(15.2) | 160 | 20 | 8.91(8.78) |

**Supplementary Table 3.** Possible combinations of alcohol solvents and good solvents followed by the maximum solvent ratio that positively charged PbSe QDs maintain colloid stable.

| Alcohol solvent | Good solvent | Maximum ratio |
|---|---|---|
| Methanol | Toluene | 1:3 |
| Ethanol | Hexane | 1:2 |
| Ethanol | Chloroform | 1:2 |
| Ethanol | Toluene | 1:1 |
| Isopropanol | Toluene | 1:2 |
| Butanol | Toluene | 1:1 |

**Supplementary Table 4.** Average roughness in the range of 16*16 um$^2$ of PbSe QDs film fabricated by EPD at different solvent ratios of methanol/toluene.

| Samples | Solvent ratio (methanol/toluene) | Thickness (μm) | Average roughness (nm) |
|---------|----------------------------------|----------------|------------------------|
| S1      | 1:5                              | 1.85           | 10                     |
| S2      | 1:4.5                            | 2.00           | 25                     |
| S3      | 1:4                              | 2.51           | 11                     |
| S4      | 1:3.5                            | 2.92           | 11                     |
| S5      | 1:3                              | 3.42           | 20                     |
| S6      | 1:2.5                            | 4.20           | 23                     |
| S7      | 1:2                              | 5.21           | 12                     |

**Supplementary Table 5.** Dynamic light scattering results of positively charged CdSe based QDs induced by alcohol.

| Samples | Diameter distribution (nm) | Particle dispersion index, PDI |
|---------|----------------------------|--------------------------------|
| native 16mg/mL       | 22.67±7.58  | 0.112 |
| methanol/toluene 1:20 | 21.9±4.453  | 0.047 |
| ethanol/toluene 1:20  | 22.47±6.509 | 0.069 |
| ethanol/toluene 1:15  | 22.47±4.745 | 0.048 |
| ethanol/toluene 1:10  | 23.03±5.786 | 0.061 |
| butanol/toluene 1:20  | 22.6±3.594  | 0.032 |

**Supplementary Table 6.** The calculation details of the ligand amount of a single PbSe QD with the mass of a single 4.64 nm PbSe QD is calibrated to 10.

| Sample diameter (nm) | Morphology | Volume (nm$^3$) | Mass (a.u.) | Mass fraction of the ligand (%) | Ligand amount (a.u.) |
|----------------------|------------|-----------------|-------------|---------------------------------|----------------------|
| 4.64  | sphere | 52.3061    | 10        | 18.0588 | 1.8059  |
| 8.35  | sphere | 304.83     | 58.2781   | 15.7059 | 9.1531  |
| 11.35 | cube   | 1462.1354  | 279.5344  | 6.7647  | 18.9097 |
| 17.22 | cube   | 5106.2191  | 976.2187  | 4.6471  | 45.3659 |
| 22.68 | cube   | 11666.1928 | 2230.3695 | 3.5882  | 80.0301 |

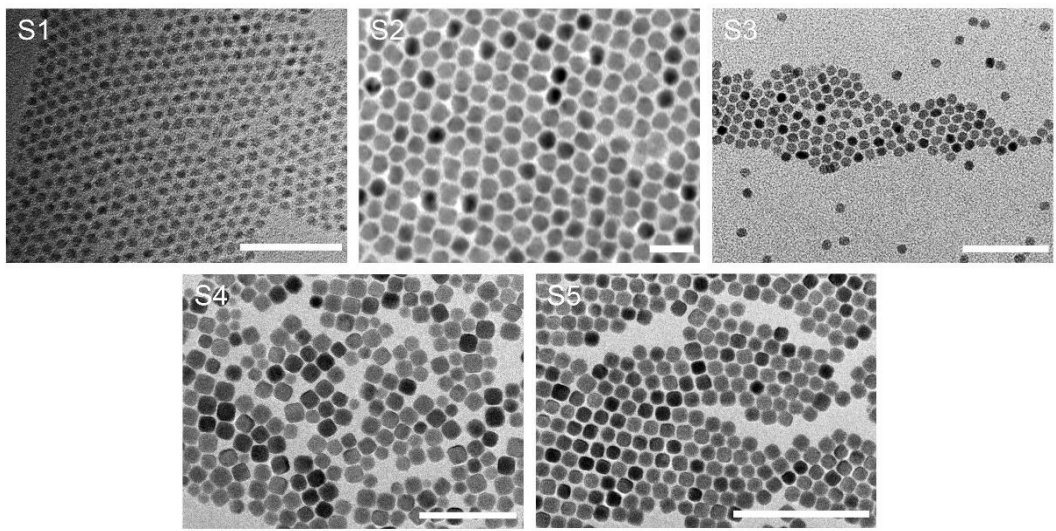

**Supplementary Figure 1.** TEM images of PbSe quantum dots.

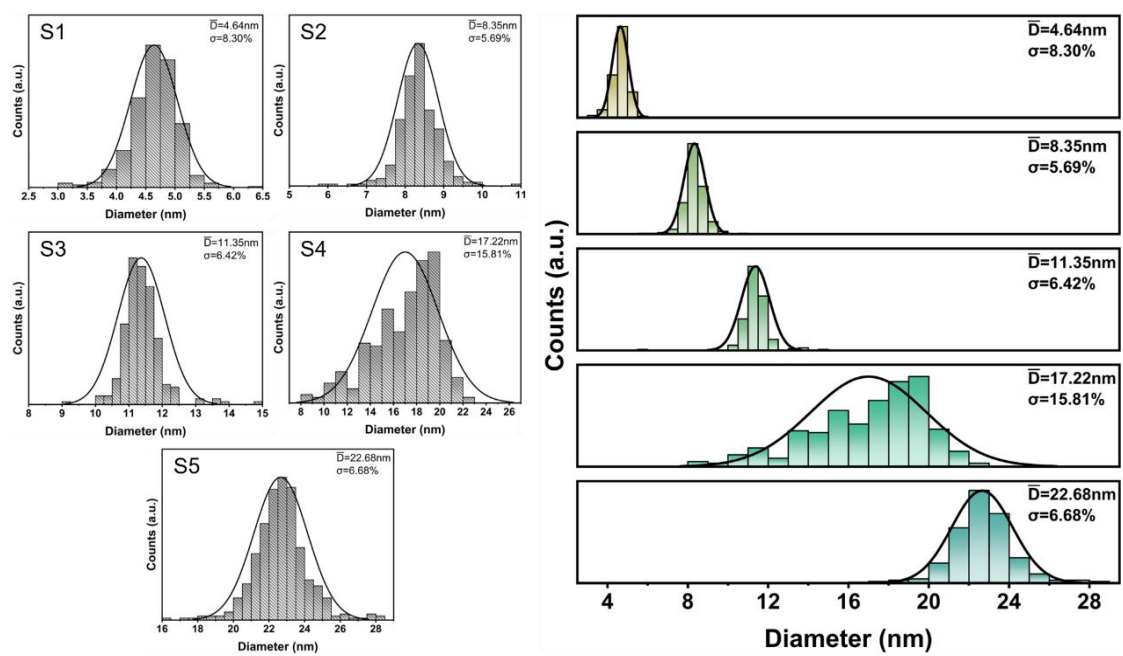

**Supplementary Figure 2.** Size distributions of PbSe quantum dots. Several hundreds of quantum dots were counted for each sample.

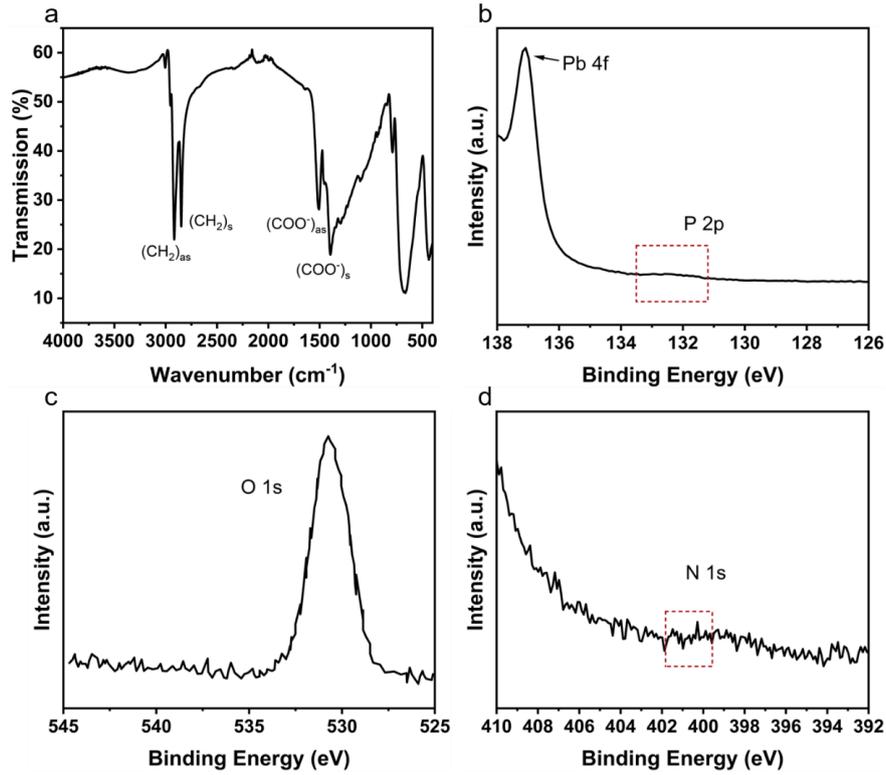

**Supplementary Figure 3**. FTIR and XPS plots of OA capped PbSe quantum dots, showing that the surface is free of OLA and TOP ligands.

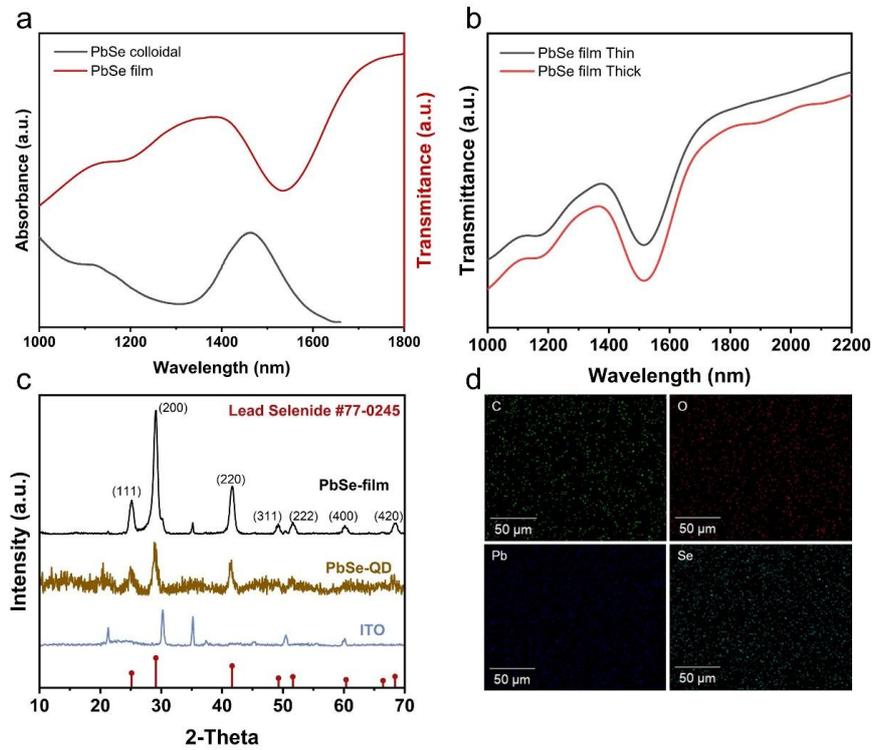

**Supplementary Figure 4.** (a), (b) Absorption and transmission spectroscopy, (c) XRD and (d) EDS mapping plots of PbSe quantum dots film fabricated by EPD

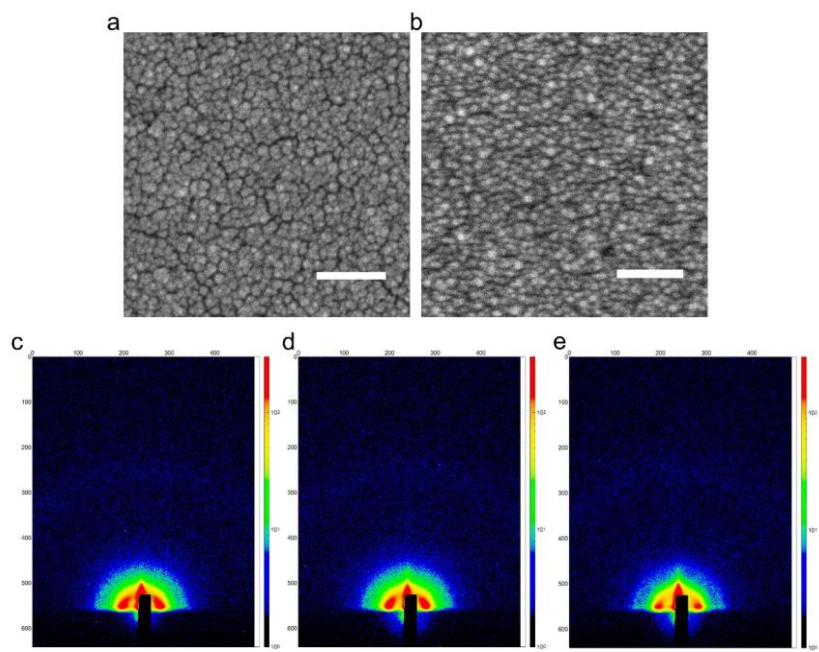

**Supplementary Figure 5**. SEM images and GISAXS plots of PbSe quantum dots film fabricated by controllable EPD. Typical SEM images of PbSe film obtained by (a) methanol/toluene and (b) ethanol/hexane solvents show a disordering arrangement and an ordering arrangement of particles. The scale bar is 100 nm. By regulating EPD parameters, films with increasingly higher particle-oriented arrangement can be fabricated as shown in (c-e) GISAXS plots.

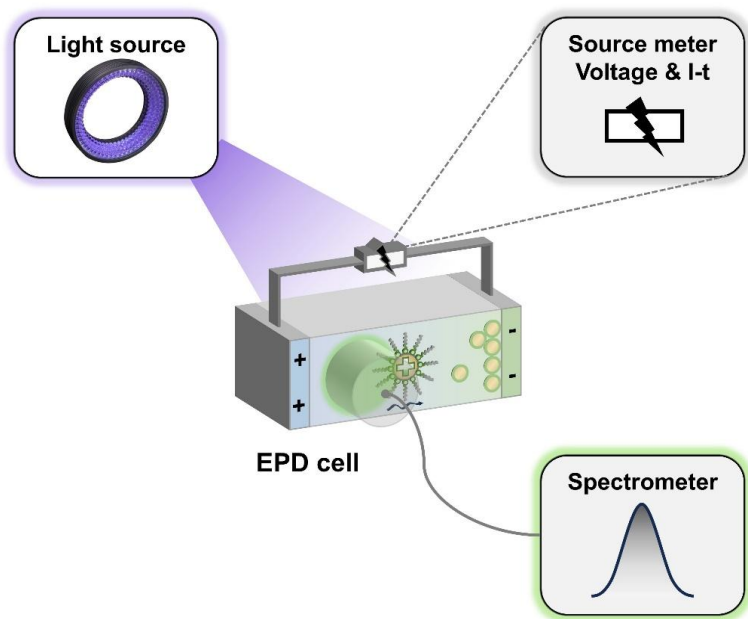

**Supplementary Figure 6**. Schematic diagrams showing the deposition flux monitoring setup combining the in-situ absorption spectra and current-time curves.

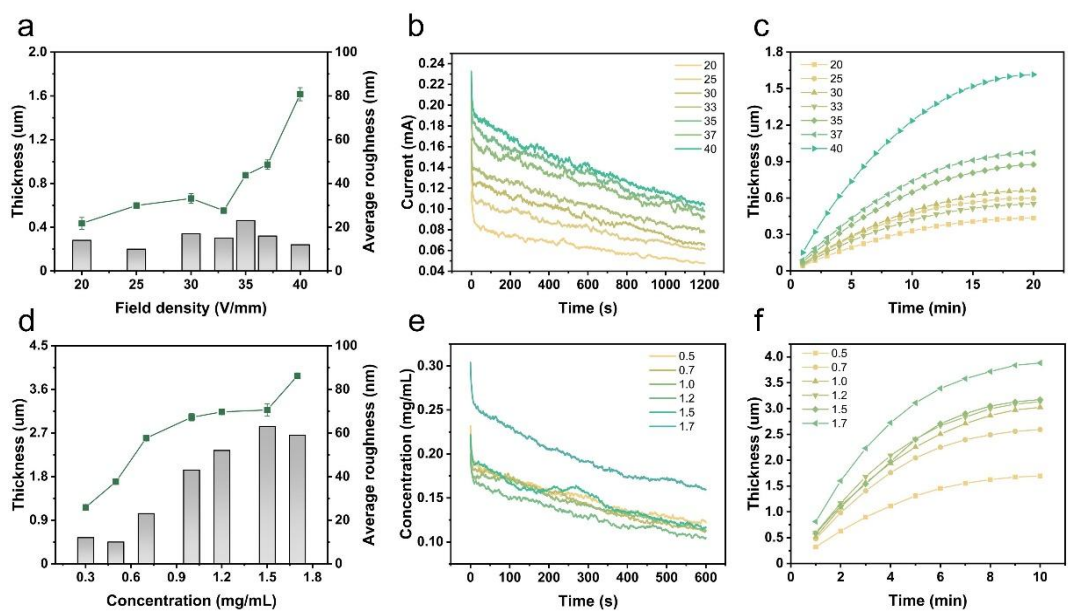

**Supplementary Figure 7**. Deposition thickness and detailed current-time curves controlled by (a-c) electric field and (d-f) quantum dots concentration. The thickness-time curves were obtained by integrating the current peak.

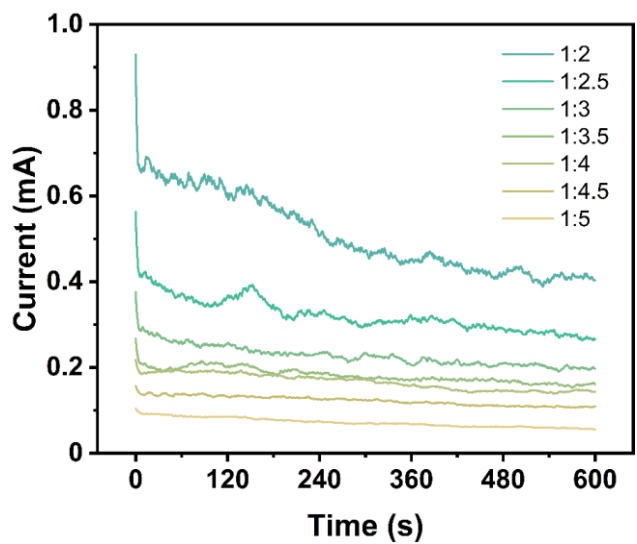

**Supplementary Figure 8**. The current-time curve during EPD of PbSe quantum dots in methanol/toluene solvent at different volume ratio, with the field density 40 V/mm and the concentration 0.7 mg/mL.

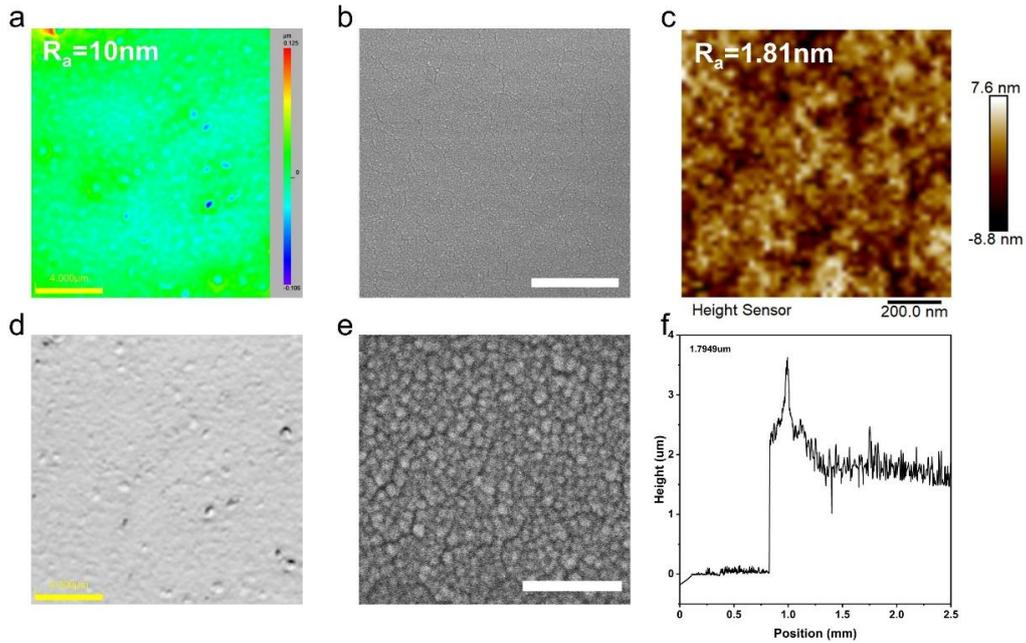

**Supplementary Figure 9**. (a) Two-dimensional height distribution image and (d) laser imaging in confocal laser scanning microscope (CLSM), (b), (e) SEM images, (c) AFM image and (f) STEP plots of a typical PbSe quantum dots film fabricated by EPD (methanol/toluene-1:5, field density 40 V/mm, concentration 0.7 mg/mL), with the bar 1um for (b) and 100nm for (e).

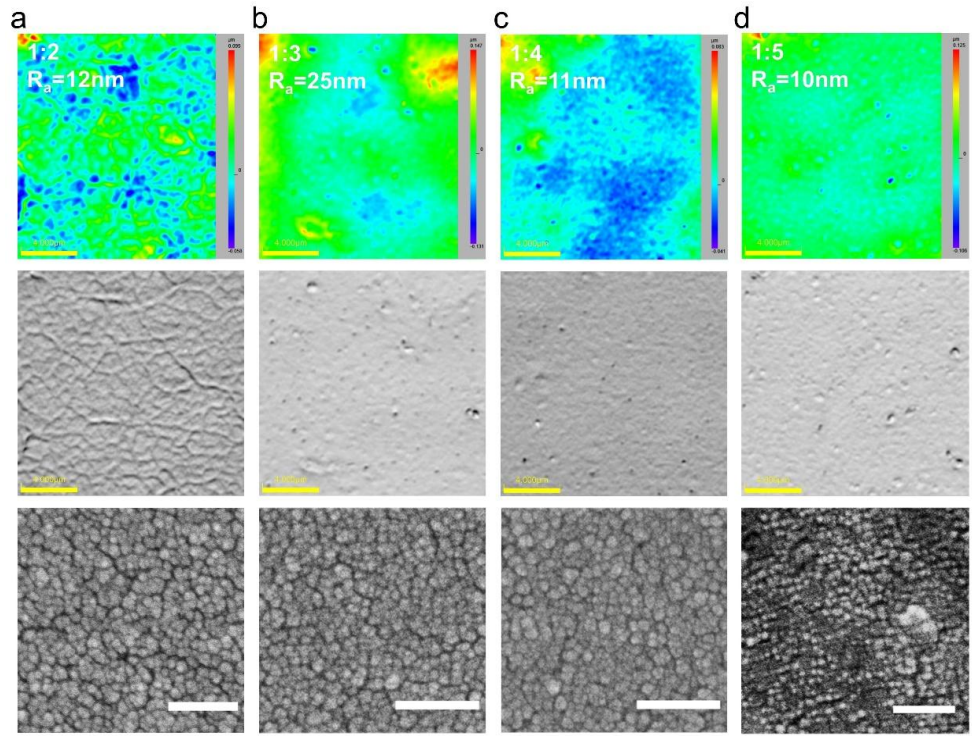

**Supplementary Figure 10**. CLSM and SEM images and the corresponding roughness of PbSe quantum dots film fabricated by EPD in methanol/toluene solvent at volume ratio (a) 1:2, (b) 1:3, (c) 1:4, (d) 1:5. The scale bar in SEM is 100 nm.

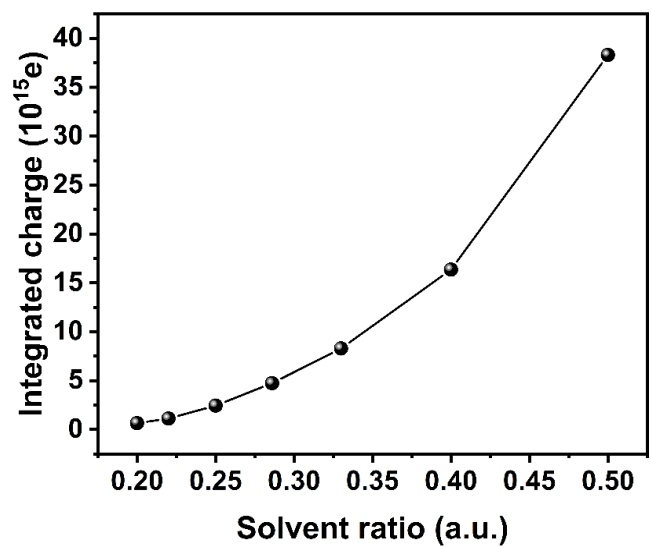

**Supplementary Figure 11**. The total charge transferred between the PbSe quantum dots and electrodes during EPD at different solvent ratio of methanol/toluene by integrating the current-time curves.

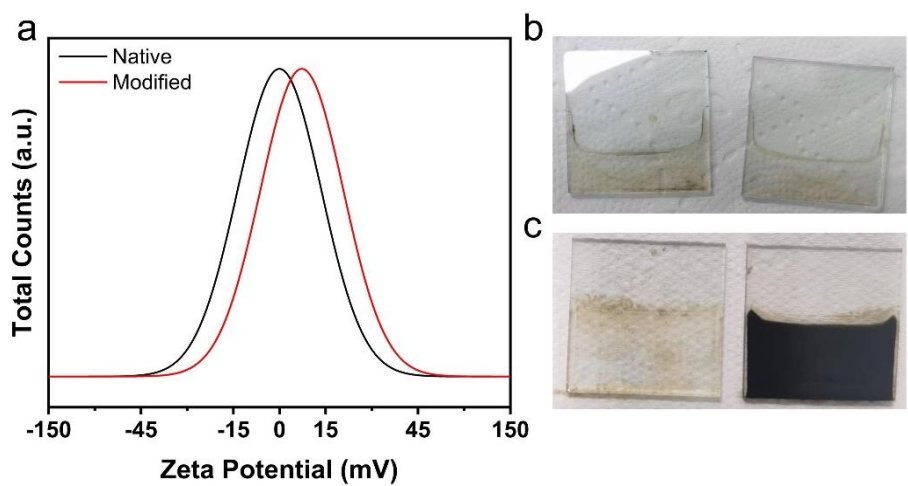

**Supplementary Figure 12**. (a) The zeta potential distribution and (b), (c) deposition results of PbSe quantum dots before and after the adding of alcohol solvent (ethanol/hexane-1:3).

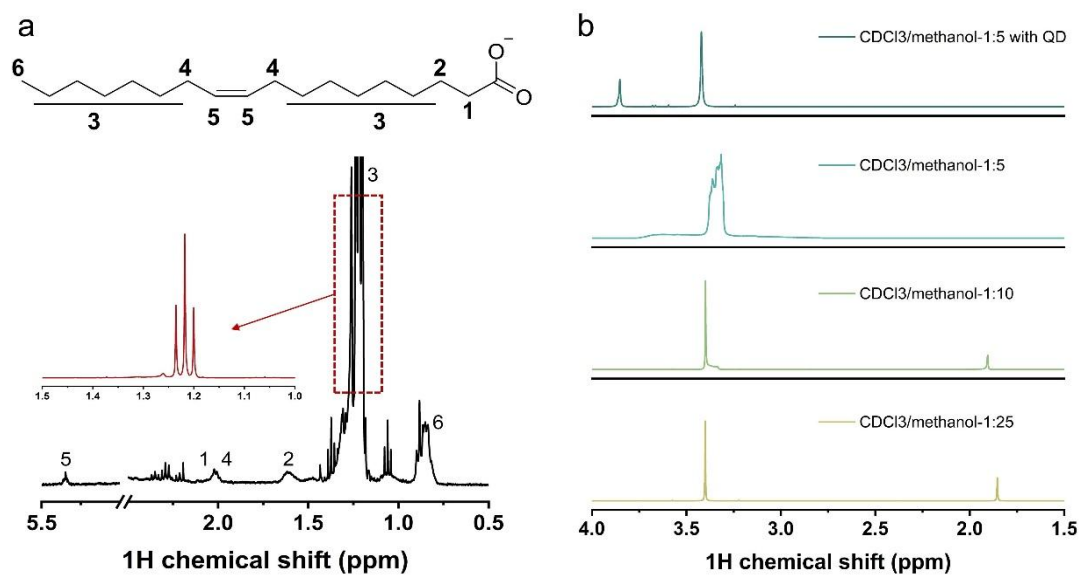

**Supplementary Figure 13**. The $^1$HNMR spectroscopy of (a) OA ligand of PbSe quantum dots, and (b) methanol at different volume ratio of methanol/CDCl$_3$.

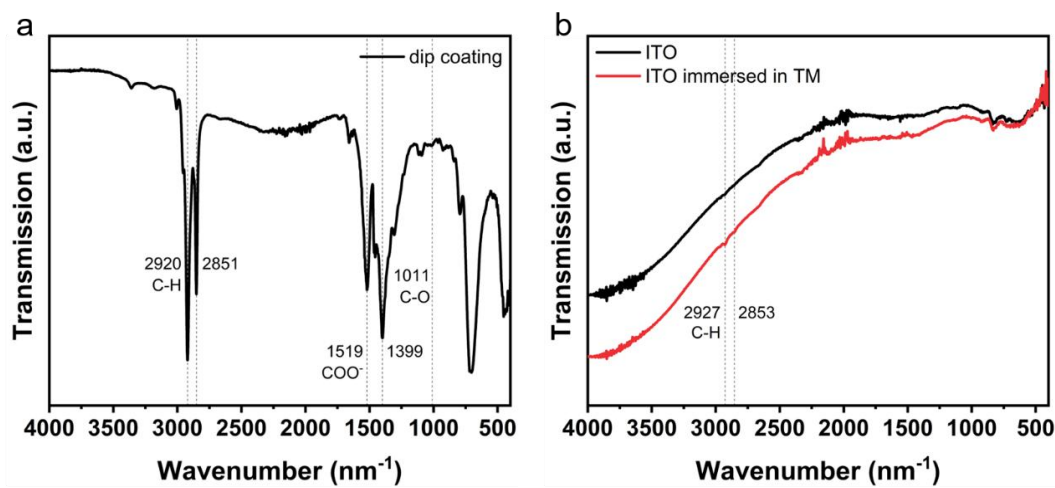

**Supplementary Figure 14**. FTIR plots of PbSe quantum dots film fabricated by (a) dip coating and (b) immersion in methanol/toluene, showing that the countercharged molecules obtained by methanol dissociation are unique to EPD process.

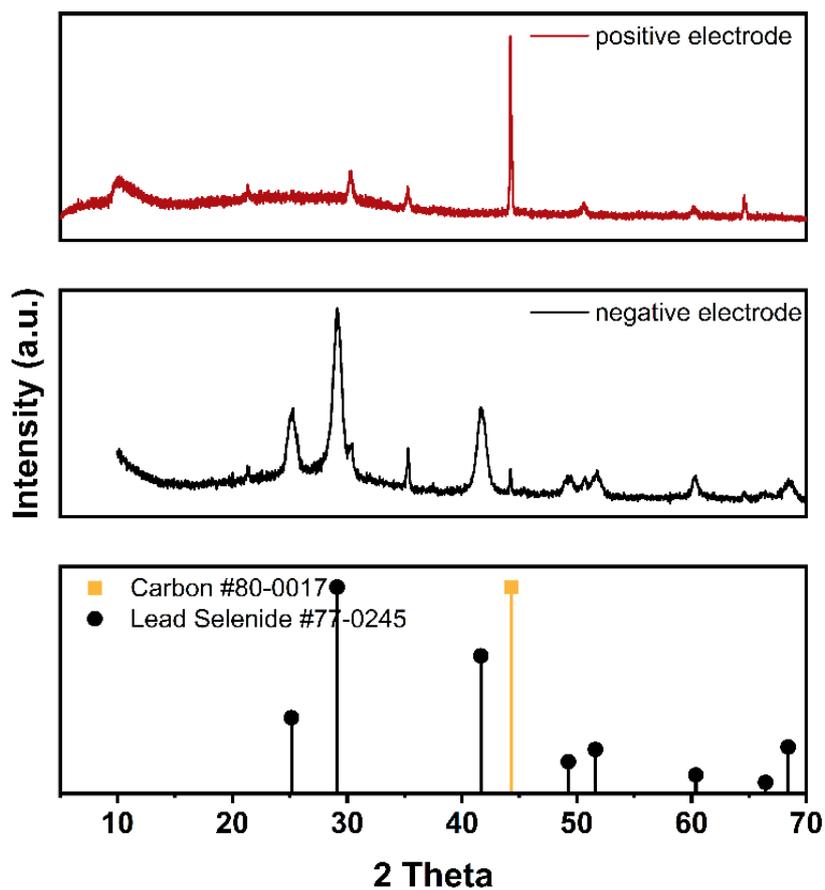

**Supplementary Figure 15**. XRD plots of deposit on electrodes resulted from EPD of PbSe quantum dots in methanol/toluene solvent.

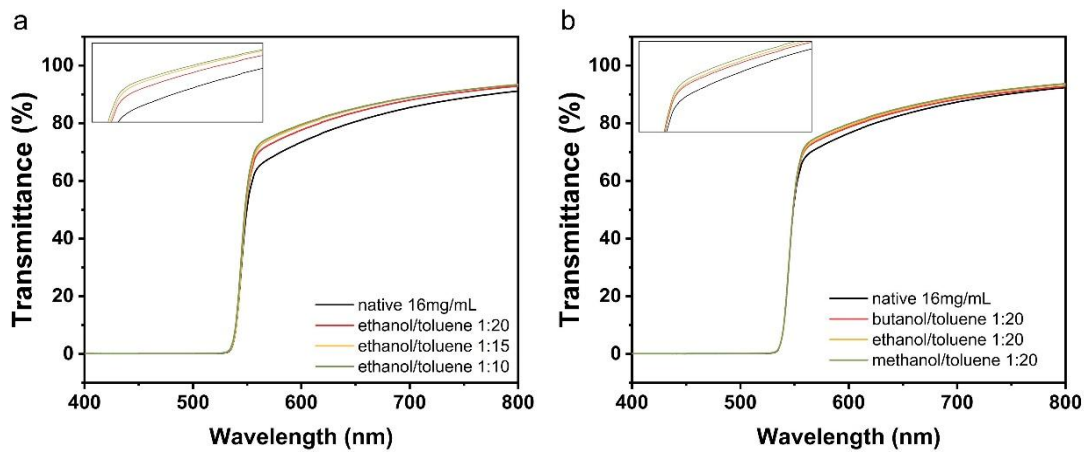

**Supplementary Figure 16**. The scattering light in the absorption spectrum of CdSe-based quantum dots in (a) different volume ratio of ethanol/toluene and (b) different kinds of alcohol.

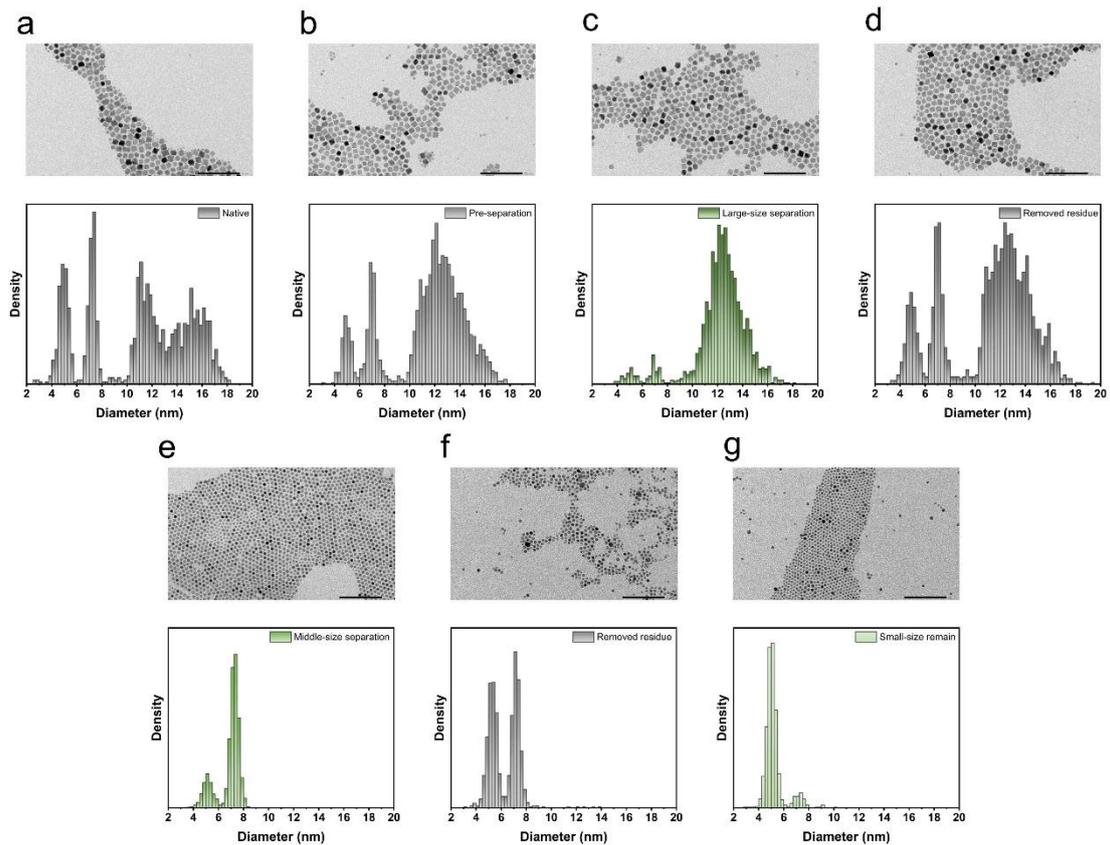

**Supplementary Figure 17**. TEM images and corresponding size distribution of (a) target sample and (b-g) separated samples with the scale bar 100 nm. About ten thousand of quantum dots were counted for each sample through a convolutional neural network-based detection and segmentation model and its supporting software.

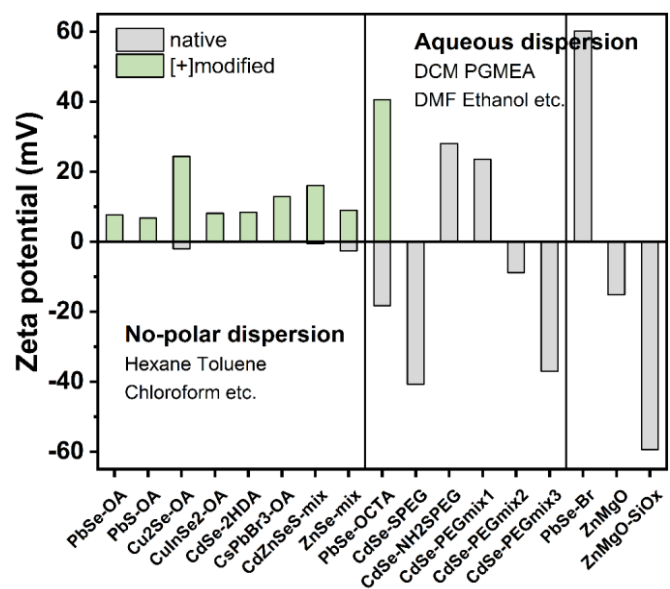

**Supplementary Figure 18**. The ethanol induced (1:4 in volume) zeta potential of carboxylic acid capped different kinds of quantum dots in non-polar solvent, and the comparison with modified quantum dots dispersed in polar solvent.

**Supplementary References**

1. Zhu, D.; Wang, L.; Liu, Z.; Tang, A. Effects of Surface Ligands on Localized Surface Plasmon Resonance and Stabilization of Cu2−xSe Nanocrystals. *Appl. Surf. Sci.* **2020**, *509*, 145327.
2. Lian, W.; Tu, D.; Weng, X.; Yang, K.; Li, F.; Huang, D.; Zhu, H.; Xie, Z.; Chen, X. Near-Infrared Nanophosphors Based on CuInSe$_2$ Quantum Dots with Near-Unity Photoluminescence Quantum Yield for Micro-LEDs Applications. *Adv. Mater.* **2024**, *36*, 2311011.
3. Tang, J.; Li, F.; Yang, G.; Ge, Y.; Li, Z.; Xia, Z.; Shen, H.; Zhong, H. Reducing the Chromaticity Shifts of Light-Emitting Diodes Using Gradient-Alloyed CdxZn1−xSeyS1−y@ZnS Core Shell Quantum Dots with Enhanced High-Temperature Photoluminescence. *Adv. Opt. Mater.* **2019**, *7*, 1801687.
4. Chen, X.; Lin, X.; Zhou, L.; Sun, X.; Li, R.; Chen, M.; Yang, Y.; Hou, W.; Wu, L.; Cao, W.; Zhang, X.; Yan, X.; Chen, S. Blue Light-Emitting Diodes Based on Colloidal Quantum Dots with Reduced Surface-Bulk Coupling. *Nat. Commun.* **2023**, *14*, 284.
5. Yang, M.; Bao, H.; Jing, Y.; Ao, Z.; Wang, J.; Wu, L.; Feng, Y.; Song, T.; Wu, X.; Yan, Y.; Zhong, H. Siloxane-Modified ZnMgO Nanoparticles with Enhanced Colloidal Stability and Improved Quantum Dot Light-Emitting Diode Stability. *J. Phys. Chem. Lett.* **2025**, *16*, 2764–2770.
6. Islam, M. A.; Herman, I. P. Electrodeposition of Patterned CdSe Nanocrystal Films Using Thermally Charged Nanocrystals. *Appl. Phys. Lett.* **2002**, *80*, 3823–3825.
7. Islam, M. A.; Xia, Y.; Steigerwald, M. L.; Yin, M.; Liu, Z.; O'Brien, S.; Levicky, R.; Herman, I. P. Addition, Suppression, and Inhibition in the Electrophoretic Deposition of Nanocrystal Mixture Films for CdSe Nanocrystals with γ-Fe$_2$O$_3$ and Au Nanocrystals. *Nano Lett.* **2003**, *3*, 1603–1606.
8. Mahajan, S. V.; Kavich, D. W.; Redigolo, M. L.; Dickerson, J. H. Structural Properties of Electrophoretically Deposited Europium Oxide Nanocrystalline Thin Films. *J. Mater. Sci.* **2006**, *41*, 8160–8165.
9. Ahmed, S.; Ryan, K. M. Centimetre Scale Assembly of Vertically Aligned and Close Packed Semiconductor Nanorods from Solution. *Chem. Commun.* **2009**, No. 42, 6421.
10. Salant, A.; Shalom, M.; Hod, I.; Faust, A.; Zaban, A.; Banin, U. Quantum Dot Sensitized Solar Cells with Improved Efficiency Prepared Using Electrophoretic Deposition. *ACS Nano*. **2010**, *4*, 5962–5968.
11. Parsi Benehkohal, N.; González-Pedro, V.; Boix, P. P.; Chavhan, S.; Tena-Zaera, R.; Demopoulos, G. P.; Mora-Seró, I. Colloidal PbS and PbSeS Quantum Dot Sensitized Solar Cells Prepared by Electrophoretic Deposition. *J. Phys. Chem. C*. **2012**, *116*, 16391–16397.
12. Song, K. W.; Costi, R.; Bulović, V. Electrophoretic Deposition of CdSe/ZnS Quantum Dots for Light-Emitting Devices. *Adv. Mater.* **2013**, *25*, 1420–1423.
13. Lhuillier, E.; Hease, P.; Ithurria, S.; Dubertret, B. Selective Electrophoretic Deposition of CdSe Nanoplatelets. *Chem. Mater.* **2014**, *26*, 4514–4520.
14. Otelaja, O. O.; Ha, D.-H.; Ly, T.; Zhang, H.; Robinson, R. D. Highly Conductive Cu$_{2-x}$S Nanoparticle Films through Room-Temperature Processing and an Order of Magnitude Enhancement of Conductivity via Electrophoretic Deposition. *ACS Appl. Mater. Interfaces*. **2014**, *6*, 18911–18920.
15. Dillon, A. D.; Le Quoc, L.; Goktas, M.; Opasanont, B.; Dastidar, S.; Mengel, S.; Baxter, J. B.; Fafarman, A. T. Thin Films of Copper Indium Selenide Fabricated with High Atom Economy by Electrophoretic Deposition of Nanocrystals under Flow. *Chem. Eng. Sci.* **2016**, *154*, 128–135.
16. Yu, Y.; Yu, D.; Orme, C. A. Reversible, Tunable, Electric-Field Driven Assembly of Silver

Nanocrystal Superlattices. *Nano Lett.* **2017**, *17*, 3862–3869.

17. Ravi, V. K.; Scheidt, R. A.; DuBose, J.; Kamat, P. V. Hierarchical Arrays of Cesium Lead Halide Perovskite Nanocrystals through Electrophoretic Deposition. *J. Am. Chem. Soc.* **2018**, *140*, 8887–8894.
18. Fulari, A. V.; Thanh Duong, N.; Anh Nguyen, D.; Jo, Y.; Cho, S.; Young Kim, D.; Shrestha, N. K.; Kim, H.; Im, H. Achieving Direct Electrophoretically Deposited Highly Stable Polymer Induced CsPbBr3 Colloidal Nanocrystal Films for High-Performance Optoelectronics. *Chem. Eng. J.* **2022**, *433*, 133809.
19. Xu, X.; Kweon, K. E.; Keuleyan, S.; Sawvel, A.; Cho, E. J.; Orme, C. Rapid In Situ Ligand-Exchange Process Used to Prepare 3D PbSe Nanocrystal Superlattice Infrared Photodetectors. *Small.* **2021**, *17*, 2101166.
20. Dillon, A. D.; Mengel, S.; Fafarman, A. T. Influence of Compact, Inorganic Surface Ligands on the Electrophoretic Deposition of Semiconductor Nanocrystals at Low Voltage. *Langmuir.* **2018**, *34*, 9598–9605.
21. Zhao, J.; Chen, L.; Li, D.; Shi, Z.; Liu, P.; Yao, Z.; Yang, H.; Zou, T.; Zhao, B.; Zhang, X.; Zhou, H.; Yang, Y.; Cao, W.; Yan, X.; Zhang, S.; Sun, X. W. Large-Area Patterning of Full-Color Quantum Dot Arrays beyond 1000 Pixels per Inch by Selective Electrophoretic Deposition. *Nat. Commun.* **2021**, *12*, 4603.